\documentclass{iopart}
\usepackage{iopams}
\usepackage{enumitem}
\usepackage{braket}
\usepackage{setstack}
\usepackage{color}
\usepackage[colorlinks=true,linkcolor=blue,citecolor=blue,urlcolor=blue,bookmarks]{hyperref}
\usepackage[compress]{cite}
\usepackage[outline]{contour}
\usepackage[utf8]{inputenc}
\usepackage[T1]{fontenc}
\usepackage{doi}
\usepackage{ifthen}
\usepackage{dsfont}
\usepackage{etoolbox}

\usepackage[top=1.25 in, bottom=1.25 in, left=1.25 in, right=1.25 in]{geometry}

\usepackage{graphicx}
\usepackage{pgfplots}
\pgfplotsset{compat=1.18} 
\usepackage{tikz}
\usetikzlibrary{arrows.meta}
\tikzset{
myptr/.style={-{Stealth[scale=0.5]}},
}
\usetikzlibrary{patterns}

\newcommand{\eqref}[1]{\eref{#1}}
\renewcommand{\bs}[1]{\boldsymbol{#1}}

\usepackage{cancel}

\def\ii{{\rm i}}
\def\xx{{\rm x}}
\def\yy{{\rm y}}
\def\zz{{\rm z}}
\def\kar{\chi}

\newtheorem{theorem}{Conjecture}

\AtBeginEnvironment{quote}{\itshape}

\begin{document}
\title{Integrability and charge transport in asymmetric quantum-circuit geometries}
\author{Chiara Paletta$^1$, Urban Duh$^1$, Bal\'azs Pozsgay$^2$, and Lenart Zadnik$^1$}
\address{$^1$ Department of Physics, Faculty of Mathematics and Physics, University of Ljubljana, Jadranska 19, Ljubljana SI-1000, Slovenia}
\address{$^2$ MTA-ELTE ``Momentum'' Integrable Quantum Dynamics Research Group, Department of Theoretical Physics, ELTE E\"otv\"os Lor\'and University, P\'azm\'any P\'eter s\'et\'any 1/A, Budapest 1117, Hungary}
\ead{lenart.zadnik@fmf.uni-lj.si}

\begin{abstract}
We revisit the integrability of quantum circuits constructed from two-qubit unitary gates $U$ that satisfy the Yang-Baxter equation. A brickwork arrangement of $U$ typically corresponds to an integrable Trotterization of some Hamiltonian dynamics. Here, we consider more general circuit geometries which include circuits without any nontrivial space periodicity. We show that any time-periodic quantum circuit in which $U$ is applied to each pair of neighbouring qubits exactly once per period remains integrable. We further generalize this framework to circuits with time-varying two-qubit gates. The spatial arrangement of gates in the integrable circuits considered herein can break the space-reflection symmetry even when $U$ itself is symmetric. By analyzing the dynamical spin susceptibility on ballistic hydrodynamic scale, we investigate how an asymmetric arrangement of gates affects the spin transport. While it induces nonzero higher odd moments in the dynamical spin susceptibility, the first moment, which corresponds to a drift in the spreading of correlations, remains zero. We explain this within a quasiparticle picture which suggests that a nonzero drift necessitates gates acting on distinct degrees of freedom.
\end{abstract}

\smallskip
\noindent
\emph{To the memory of Marko Medenjak, a dear friend and a brilliant colleague.}
\smallskip

\section{Introduction}

Integrable models have been crucial in understanding various statistical mechanics and condensed matter physics phenomena, ranging from phase transitions~\cite{baxter1982} to out-of-equilibrium dynamics of extended quantum systems~\cite{calabrese2016}. Most of exactly-solvable one-dimensional (1D) models in quantum many-body physics are integrable within the Yang-Baxter formalism. Here, the key role is played by the so-called $R$ matrix. This is a solution of the Yang-Baxter equation, which ensures consistent factorization of scattering between long-lived quasiparticles stabilized by abundant conserved integrals of motion. 

The $R$ matrix is the fundamental building block of integrable Hamiltonians and the algebraic Bethe ansatz method for their diagonalization~\cite{faddeev1996}. $R$ matrices can also be used in the construction of integrable discrete-time dynamical systems and Floquet protocols~\cite{destri1987,faddeev1994,gritsev2017,vanicat2018,vanicat2018integrable}. In particular, by using them as gates arranged into a quantum circuit, one can construct integrable dynamical systems that serve as time-discretizations of integrable Hamiltonian models~\cite{vanicat2018}. Such discrete-spacetime models have a dual advantage over continuous-time models. Firstly, in the context of classical simulation of the dynamics they eliminate the need for the computationally expensive ``Trotter-Suzuki limit''~\cite{suzukiGeneralizedTrottersFormula1976}, which requires an extensive number of discrete steps to reproduce a Hamiltonian evolution. Secondly, such dynamical systems naturally fit within the architecture of modern quantum computing platforms, since they correspond to sequences of more or less elementary quantum operations~\cite{kauffman2004,zhang2005,zhang2024geometric,zhang2024optimal}. 
These advantages make integrable quantum circuits a convenient platform for the investigation of spin-transport phenomena~\cite{ljubotinaBallisticSpinTransport2019,
ljubotinaKardarparisizhangPhysicsQuantum2019,rosenbergDynamicsMagnetizationInfinite2024,summer2024}, as well as good candidates for benchmarking quantum devices~\cite{aleiner2021,maruyoshi2023,hillberry2024,hutsalyuk2024}. Apart from such practical advantages, Yang-Baxter-integrable quantum circuits may exhibit exceptional stability under perturbations~\cite{morvan2022,hudomal2024,surace2024}, unconventional symmetries~\cite{znidaricInhomogeneousSU2Symmetries2024,vernier2024}, and a phase structure that is somewhat richer than in their continuous-time counterparts~\cite{ljubotinaBallisticSpinTransport2019,vernier2023,znidaricIntegrabilityGenericHomogeneous2024,suchsland2025}. Moreover, they have played an important role in the analysis of spectral form factors in certain chaotic models~\cite{friedman2019}, and in the studies of the dissipative dynamics~\cite{vanicat2018,sa2021,paletta2025,popkov2025}. 
Recently, the concept of integrable discrete-spacetime dynamical systems has also been explored in higher dimensions and further developed in the context of superintegrability~\cite{gombor2022,gombor2024,singh2024,sinha2024}.

Similar to continuous-time integrable models, the large-scale dynamics of Yang-Baxter-integrable quantum circuits can be described within the framework of generalized hydrodynamics~\cite{castro-alvaredoEmergentHydrodynamicsIntegrable2016,
bertiniTransportOutofequilibriumXXZ2016,doyon2020}. In discrete-time integrable models, generalized hydrodynamics reveals features that do not necessarily appear in their continuous-time counterparts. These features include macroscopic effects of localized perturbations~\cite{hubner2025} (see Refs.~\cite{zauner2015,bertini2016,eisler2020,gruber2021,bidzhiev2022,fagotti2022,zadnik2022} for exceptional cases of the latter in continuous-time dynamics), and spatially asymmetric dynamical correlation functions~\cite{zadnikQuantumManybodySpin2024a,richelli2024}. A common feature of quantum circuits exhibiting asymmetric spreading of correlations is the presence of gates that explicitly break the space-reflection symmetry by acting on distinct degrees of freedom~\cite{gong2022}. If the gates are symmetric, the space-reflection symmetry can still be broken by their arrangement inside the circuit. This motivates the investigation of integrable quantum circuits in which gates are not arranged in the standard brickwork fashion---such circuits have been identified and classified according to their spectral characteristics and space-time symmetries in Ref.~\cite{duhClassificationSamegateQuantum2024}. 

In this work, we investigate them systematically: we explicitly demonstrate their integrability and place them within the class of Yang-Baxter integrable physical models. We explain the main features of their classification, put forward in Ref.~\cite{duhClassificationSamegateQuantum2024}, using the Bethe ansatz equations which determine the circuits' spectra. Furthermore, we investigate the thermodynamic states of integrable circuits and derive the ingredients for the hydrodynamic description of the dynamical spin susceptibility. This allows us to investigate how the circuit geometry affects the spin transport in such discrete-spacetime dynamical systems. 

First and foremost, our findings generalize the observation of Ref.~\cite{znidaricIntegrabilityGenericHomogeneous2024} that all homogeneous brickwork circuits composed of U(1)-invariant nearest-neighbour unitary gates are integrable (see also Refs.~\cite{sogo1982,beisert2013,vieira2018,deLeeuw2020,yangbaxterboostdeleeuw} for integrability of U(1)-conserving Hamiltonians). In particular, from our observations, it follows that
\begin{quote}
    all quantum unitary circuits in which a fixed U(1)-invariant gate is applied to each pair of neighbouring qubits exactly once per period are integrable.
\end{quote}
In fact, this statement could be generalized to all gates that solve the Yang-Baxter equation. For clarity, however, we will focus on the simplest example---the isotropic Heisenberg gate. We should mention that our construction of integrable circuit geometries partially overlaps with the work of Miao et al.~\cite{miaoFloquetBaxterisation2024}. In it, the Authors describe how to construct integrable Floquet circuits in which the space and time periods are the same, and the system length is their multiple. Importantly, in our construction, there is no such constraint on the system size.

Furthermore, by investigating the hydrodynamic mode decomposition of the dynamical spin susceptibility, we demonstrate that an asymmetric arrangement of spatially-symmetric gates leads to the asymmetry of dynamical correlation functions. However, we find that it only only affects the higher-order odd moments of the dynamical spin susceptibility. The first (non-centered) moment, which corresponds to a drift (i.e., a background velocity) in the spreading of correlations, still remains zero. Our findings thus suggest that 
\begin{quote}
    the presence of background drift in dynamical correlation functions requires gates acting on pairs of distinct degrees of freedom.
\end{quote}
Below, we provide more background and outline the structure of the paper.

\subsection{Background and outline}
\label{sec:integrable-geometries}

The general setting of this work are integrable time-periodic quantum circuits composed of two-qubit nearest-neighbour gates. The circuits that we will consider generalize the brickwork circuit geometry, whose propagator for one period in time can be represented as:
\begin{equation}
        \begin{tikzpicture}[baseline=(current  bounding  box.center),scale=1]
        \foreach \x in {1,...,10}
        {
        \draw[black, line width=0.3mm] (0.5*\x,-1) -- (0.5*\x,0.75);
        }
        \node[anchor=north] at (0.5,-1){\footnotesize{1}};
        \node[anchor=north] at (1,-1){\footnotesize{2}};
        \node[anchor=north] at (1.5,-1){\footnotesize{3}};
        \node[anchor=north] at (2,-1){\footnotesize{4}};
        \node[anchor=north] at (4.5,-1){\footnotesize{$L\!-\!1$}};
        \node[anchor=north] at (5,-1){\footnotesize{$L$}};
        \foreach \x in {0,...,5}
        {
        \draw[fill=red!10!green!50!blue!20, opacity=1, rounded corners = 2,thick] (-0.125+\x,0) rectangle ++(0.75,0.5);
        \draw[fill=red!10!green!50!blue!20, opacity=1, rounded corners = 2,thick] (-0.625+\x,-0.75) rectangle ++(0.75,0.5);
        }
        \fill[white] (-0.75,-1.25) rectangle ++(1,2);
        \fill[white] (5.25,-1.25) rectangle ++(1,2);
        \fill[white] (2.75,-1.25) rectangle ++(1,2);
        \node[anchor=center] at (3.25,-0.125){$\ldots$};
        \end{tikzpicture}
        \label{eq:brickwork-geometry}
\end{equation}
Here, vertical lines signify the evolution of qubits, quantum unitary gates $U$ are represented as green rectangles applied from bottom to top, and periodic boundary conditions are assumed---Pauli matrices acting on qubits satisfy $\sigma_{L+j}^\alpha=\sigma^\alpha_j$ for $j\in\{1,\ldots,L\}$ and $\alpha\in\{\xx,\yy,\zz\}$, where $L$ is the system size, i.e., the number of qubits. 

We recall the Yang-Baxter formalism and the construction of a brickwork-type integrable Trotterization in \textbf{Section~\ref{sec:brickwork}}. By integrable Trotterization we refer to an integrable quantum circuit which, in a particular limit, reproduces some integrable continuous-time dynamics. Section~\ref{sec:brickwork} in addition reviews a diagrammatic technique inspired by Ref.~\cite{difrancesco2006}, by means of which one can identify transfer matrices commuting with integrable circuits---this technique is used later on.

\textbf{Section~\ref{sec:gen_geometries}} is motivated by the findings of Ref.~\cite{duhClassificationSamegateQuantum2024}. The latter classifies all circuits in which a fixed two-site gate $U$ is applied on all nearest-neighbour pairs of qubits exactly once per period. Moreover, it provides indications that such circuits are integrable if $U$ is chosen properly. Here, we prove this observation. In subsection~\ref{sec:classification-L10} we further explore it case-by-case. 

More specifically, in Ref.~\cite{duhClassificationSamegateQuantum2024} it was shown that, for a given system size $L$, there always exist at most $L-1$ circuits
with different spectra (quasienergies): all circuits can be sorted in at most $L-1$ \emph{spectral equivalence classes} (SECs). Two ways of labeling these SECs were used: 

\begin{enumerate}
\item[(1)] One way is to assign to each circuit in a SEC a pair of integers $(q,r)$. Here, $q$ is the number of layers in a period, and each successive layer is a translation of the previous one by $r$ sites towards the east (right). In addition, the first layer consists of a gate that acts on the first two qubits, and which is then translated repeatedly by $q$ sites, until it wraps around the chain. For example, the brickwork circuit in Eq.~\eqref{eq:brickwork-geometry} has $(q,r)=(2,1)$. 

\item[(2)] Another, equivalent but less obvious way, was by realizing that each SEC contains a circuit whose time-period consists of two opposite staircase-arrangements of gates of respective lengths $p$ and $L-p$.
Here, at a fixed circuit length $L$, only one number is required for the classifcation: the length of the left staircase, $p$. For example, the brickwork circuit in Eq.~\eqref{eq:brickwork-geometry} is unitarily equivalent to
\begin{equation}
    \begin{tikzpicture}[baseline=(current  bounding  box.center),scale=1]
    \foreach \x in {1,2,4,5,6,7,8}
    {
    \draw[black, line width=0.3mm] (0.5*\x,-3.25) -- (0.5*\x,0.75);
    }
    \node[anchor=north] at (0.5,-3.25){\footnotesize{1}};
    \node[anchor=north] at (1,-3.25){\footnotesize{2}};
    \node[anchor=north] at (2,-3.25){\footnotesize{$\frac{L}{2}$}};
    \node[anchor=north] at (2.5,-3.25){\footnotesize{$\frac{L}{2}\!+\!1$}};
    \node[anchor=north] at (3.5,-3.25){\footnotesize{$L\!-\!1$}};
    \node[anchor=north] at (4,-3.25){\footnotesize{$L$}};
    \foreach \x in {0,1,2,3,4}
        {
        \draw[fill=red!10!green!50!blue!20, opacity=1, rounded corners = 2,thick] (-0.125+0.5*\x,-0.75*\x) rectangle ++(0.75,0.5);
        }
    \foreach \x in {5,6,7,8}
        {
        \draw[fill=red!10!green!50!blue!20, opacity=1, rounded corners = 2,thick] (-0.125+0.5*\x,-0.75*8+0.75*\x) rectangle ++(0.75,0.5);
        }
    \fill[white] (-0.75,-1.25) rectangle ++(1,2);
    \fill[white] (4.25,-1.25) rectangle ++(1,2);
    \fill[white] (1.25,-3.25) rectangle ++(0.5,4);
    \fill[white] (2.75,-3.25) rectangle ++(0.5,4);
    \node[anchor=center] at (1.5,-0.125){$\ldots$};
    \node[anchor=center] at (3,-0.125){$\ldots$};
    \end{tikzpicture}
\end{equation}
with $p=\frac{L}{2}$ being the length of the left staircase.
\end{enumerate}
We will explicitly show that, for a proper choice of $U$, all circuit geometries classified in Ref.~\cite{duhClassificationSamegateQuantum2024} are integrable. Importantly, our proof holds irrespective of whether the circuits are periodic in space, or not. The crucial role is played by quantum gates that solve the Yang-Baxter equation---a relation which underpins the integrable structure of the majority of known integrable models in physics.

Integrability of various arrangements of quantum gates might not be surprising in light of Baxter's work on solvable two-dimensional eight-vertex models of equilibrium statistical mechanics~\cite{baxter97}. Our case of quantum circuits, in which any neighbouring pair of qubits is acted upon exactly once per period, may be regarded as a non-equilibrium analog of such lattice models.  Here, we establish a clear correspondence between the selection of inhomogeneities in the transfer matrix and the geometric configuration of the gates. As a proof of concept, we dedicate subsection~\ref{sec:classification-L10} to circuits on a fixed number of qubits, specifically, to those that were depicted in Fig.~4 of Ref.~\cite{duhClassificationSamegateQuantum2024}. Our findings explain how the SEC classification number $p$ relates to the integrable structure, and why it uniquely specifies a particular class of circuits with the same spectrum of quasienergies. In subsection~\ref{sec:different-layers} we go beyond the same-gate circuits by showing how to construct integrable circuits in which subsequent layers of gates are different. These circuits contrast with the recently proposed ``fishnet circuits''~\cite{krajnik2024} in two key ways. First, fishnet circuits are defined on a lattice tilted relatively to the brickwork circuit. Second, they allow for a completely random temporal sequence of gate parameters. In contrast, in our construction, which generalizes the brickwork architecture, the gate parameters in different layers must satisfy a constraint---they are the differences of inhomogeneities in the transfer matrix, and only the latter can be assigned fixed random values. Lastly, in subsection~\ref{sec:lattice-cut} we show an alternative diagrammatic method of constructing integrable circuits with equal space- and time-periodicities.

\textbf{Section~\ref{sec:eigenstates}} follows Ref.~\cite{zadnikQuantumManybodySpin2024a} in describing Bethe ansatz diagonalization of integrable quantum circuits that are periodic in both space and time. Bethe-ansatz equations are reported in subsection~\ref{sec:bethe-eqs}: they clarify why the SEC classification number $p$ determines the spectrum. We present the thermodynamic Bethe-ansatz equations in subsection~\ref{sec:thermodynamic-content}, where we also describe the simplest thermodynamic state of our Floquet-driven systems: an infinite-temperature grand-canonical Gibbs ensemble at a finite chemical potential\footnote{A finite temperature state is not well defined, since Floquet-driven systems have no Hamiltonian.}.

In \textbf{Section~\ref{sec:hydro}} we use the thermodynamic Bethe ansatz results to investigate the transport of the U(1) charge---the total magnetization ${\rm S}^\zz$. In particular, we examine the long-time behaviour of the dynamical spin susceptibility $\langle {\rm S}^\zz(x,t){\rm S}^\zz(0,0)\rangle$. Its hydrodynamic mode decomposition relates the transport coefficients to the thermodynamic properties of the state.

The breaking of the space-reflection symmetry on the level of quantum gates has been linked to the appearance of a drift in the dynamics, for example, in the dynamical correlation functions~\cite{zadnikQuantumManybodySpin2024a,richelli2024}, and earlier, in the spreading of the entanglement~\cite{gong2022}. Some of the integrable space-and-time-periodic circuits considered herein break the space-reflection symmetry due to an asymmetric arrangement of the gates. We will show that the latter is not sufficient for the drift to appear: a drift necessarily requires quantum gates that act on different degrees of freedom (e.g., different spins). However, the asymmetry of the circuit still results in nonzero higher odd moments of the dynamical spin susceptibility. The results of this section are corroborated by extensive tensor-network simulations.

We conclude with final remarks in \textbf{Section~\ref{sec:conclusion}}.

\section{Recalling the Yang-Baxter-integrable brickwork circuits}
\label{sec:brickwork}

\subsection{Yang-Baxter equation and transfer matrices}

Integrability of the circuits which we will consider, and which were classified according to their spectral characteristics in Ref.~\cite{duhClassificationSamegateQuantum2024}, relies on the solution of the Yang-Baxter equation (YBE)
\begin{equation}
    R_{1,2}(\lambda-\eta)R_{1,3}(\lambda)R_{2,3}(\eta)=R_{2,3}(\eta)R_{1,3}(\lambda)R_{1,2}(\lambda-\eta)
    \label{eq:yang-baxter}
\end{equation} 
defined on the Hilbert space $(\mathbb{C}^2)^{\otimes 3}$ of three qubits. The solution, $R_{i,j}(\lambda)$, is a $4\times 4$ matrix which acts on two qubits in sites $i$ and $j$,  and which is of the difference form---it depends only on one spectral parameter $\lambda\in\mathbb{C}$. We focus on a regular $R$ matrix, which reduces to a permutation of two qubits at $\lambda=0$, i.e., 
\begin{equation}
    R_{i,j}(0)=P_{i,j}\equiv\frac{1}{2}(\mathds{1}+\bs{\sigma}_i\cdot\bs{\sigma}_j),
    \label{eq:shift-point}
\end{equation}
and which satisfies
\begin{equation}
    R_{i,j}(\lambda)R_{j,i}(-\lambda)=\mathds{1}.
    \label{eq:unitarity}
\end{equation}
Here, $\mathds{1}$ denotes the identity matrix, and $\bs{\sigma}=(\sigma^\xx,\sigma^\yy,\sigma^\zz)$ is the vector of Pauli matrices. In this work, we will use the SU(2)-invariant $R$ matrix of the isotropic Heisenberg model,
\begin{equation}
    R_{i,j}(\lambda)=\frac{1}{1-\ii \lambda}(P_{i,j}-\ii \lambda \mathds{1}),
    \label{eq:R-matrix}
\end{equation}
which was used in Ref.~\cite{vanicat2018} to build an integrable discrete-time version of the model. However, our construction is more general: one can build a circuit out of any other invertible solution of Eq.~\eqref{eq:yang-baxter}~\cite{ljubotinaBallisticSpinTransport2019,aleiner2021,miaoFloquetBaxterisation2024,hutsalyuk2024}. In fact, the above conditions on the $R$ matrices which can be used to construct integrable quantum circuits are surprisingly not too restrictive: any U(1)-invariant two-qubit unitary gate can be related to an $R$ matrix satisfying them~\cite{znidaricIntegrabilityGenericHomogeneous2024}, and one can even generalize the construction to systems involving different local degrees of freedom (spins)~\cite{zadnikQuantumManybodySpin2024a}, or a different number thereof~\cite{sa2021,gomborIntegrableSpinChains2021}. 

First and foremost, the $R$ matrix~\eqref{eq:R-matrix} is the essential ingredient of the algebraic Bethe ansatz---a procedure of simultaneously diagonalizing a family of transfer matrices
\begin{equation}
    \mathds{T}(\lambda)={\rm Tr}_a \left[\prod_{1\le j\le L}^\rightarrow R_{j,a}(\lambda-\nu_j)\right],
    \label{eq:transfer-matrix}
\end{equation}
which, for a fixed set of inhomogeneities $\bs{\nu}=(\nu_1,\ldots, \nu_L)$, mutually commute due to YBE~\eqref{eq:yang-baxter}:
\begin{equation}
    [\mathds{T}(\lambda),\mathds{T}(\mu)]=0.
\end{equation}
The partial trace in Eq.~\eqref{eq:transfer-matrix} is over the ancillary degree of freedom, i.e., over an auxiliary qubit denoted by index $a$. The arrow above the product indicates the direction in which index $j$ increases. 

Now, a specific choice of the inhomogeneities $\bs{\nu}$ yields a transfer matrix from which the time-evolution operator can be constructed. Diagonalizing transfer matrices thus gives access to the quasienergies and eigenstates of some time-evolution operator. Below we review how one can obtain a discrete-time brickwork-type evolution operator in the case of the $R$ matrix~\eqref{eq:R-matrix} of the isotropic Heisenberg model.

\subsection{Integrable Trotterization of the Heisenberg model} 

The $R$ matrix~\eqref{eq:R-matrix} is symmetric in the sense that $R_{i,j}(\lambda)=R_{j,i}(\lambda)$, and for $\lambda\in\mathbb{R}$ it satisfies $R_{i,j}(-\lambda)=R_{i,j}(\lambda)^\dagger$. By Eq.~\eqref{eq:unitarity} one could then take the $R$ matrix acting on two neighbouring qubits as a local unitary gate. However, we will start differently: a special role is played by the circuits which, in a certain limit, reproduce an integrable continuous-time dynamics. In such circuits, the two-qubit gate is chosen as 
\begin{equation}
    U_{j,j+1}=P_{j,j+1}R_{j,j+1}(\tau),   
    \label{eq:unitary-gate}
\end{equation}
where $\tau$ is a real gate parameter. Firstly, this is convenient because, for small $\tau$, and using the $R$ matrix~\eqref{eq:R-matrix} of the Heisenberg model as an example, we have
\begin{equation}
    U_{j,j+1}=\mathds{1}-\ii \tau P_{j,j+1}+O(\tau^2).    
\end{equation}
In the first order in $\tau$, the gate therefore corresponds to an exponential of the local Hamiltonian density $h_{j,j+1}= P_{j,j+1}\equiv\frac{1}{2}(\mathds{1}+\bs{\sigma}_j\cdot\bs{\sigma}_{j+1})$ of the integrable isotropic Heisenberg model, in accordance with the definition of the time-evolution operator for Hamiltonian systems. Secondly, we observe that the isotropic Heisenberg gate additionally satisfies $U_{j,j+1}=R_{j,j+1}(-\frac{1}{\tau})$. Hence, a circuit using $R$ matrices~\eqref{eq:R-matrix} as gates is included among the circuits built out of the gate~\eqref{eq:unitary-gate}.

The gate~\eqref{eq:unitary-gate} satisfies a slightly different version of the Yang-Baxter equation~\eqref{eq:yang-baxter}, namely
\begin{equation}
    U_{1,2}R_{1,3}(\lambda_{\tau_1})R_{2,3}(\lambda_{\tau_2})=R_{1,3}(\lambda_{\tau_2})R_{2,3}(\lambda_{\tau_1})U_{1,2},
    \label{eq:gate-RLL}
\end{equation}
where $\lambda_{\tau_{1,2}}\equiv\lambda-\tau_{1,2}$ and $\tau\equiv\tau_2-\tau_1$. Its diagrammatic representation
\begin{equation}
\begin{tikzpicture}[baseline=(current  bounding  box.center),scale=1.25]
\draw[black,line width=0.4mm] (-0.5,0) -- (1.25,0);
\node[anchor=north] at (0,-0.525){\footnotesize{$1$}};
\node[anchor=north] at (0.75,-0.525){\footnotesize{$2$}};
\node[anchor=east] at (-0.5,0){\footnotesize{$3$}};
\draw[fill=red!10!green!50!blue!20, opacity=1, rounded corners = 2,thick] (-0.125,0.5) rectangle ++(1,0.5);
\node[anchor=center] at (0.375,0.75){$\tau$};
\draw[green!20!blue, line width=1mm, opacity=0.5] (0,1.25) -- (0,1);
\draw[green!20!blue, line width=1mm, opacity=0.5] (0.75,0.5) -- (0.75,0);
\draw[red!80!blue, line width=0.5mm, opacity=0.5] (0,0.5) -- (0,0);
\draw[red!80!blue, line width=0.5mm, opacity=0.5] (0.75,1.25) -- (0.75,1);
\draw[red!80!blue, line width=0.5mm, opacity=0.5] (0,-0.25) -- (0,-0.5);
\draw[green!20!blue, line width=1mm, opacity=0.5] (0.75,-0.25) -- (0.75,-0.5);

\foreach \x in {0.5}
{
\draw[fill=white, opacity=1, rounded corners = 0.5,thick] (\x+0.25,-0.3) -- (\x+0.25+0.3,0) -- (\x+0.25,0.3) -- (\x+0.25-0.3,0) -- (\x+0.25,-0.3);
\draw[fill=green!20!blue, opacity=0.25, rounded corners = 0.5,thick] (\x+0.25,-0.3) -- (\x+0.25+0.3,0) -- (\x+0.25,0.3) -- (\x+0.25-0.3,0) -- (\x+0.25,-0.3);
\node[anchor=center] at (\x+0.25,0){\footnotesize{$\lambda_{\tau_2}$}};

\draw[fill=white, opacity=1, rounded corners = 0.5,thick] (\x-0.5,-0.3) -- (\x-0.5+0.3,0) -- (\x-0.5,0.3) -- (\x-0.5-0.3,0) -- 
(\x-0.5,-0.3);
\draw[fill=red!80!blue, opacity=0.25, rounded corners = 0.5,thick] (\x-0.5,-0.3) -- (\x-0.5+0.3,0) -- (\x-0.5,0.3) -- (\x-0.5-0.3,0) -- (\x-0.5,-0.3);
\node[anchor=center] at (\x-0.5,0){\footnotesize{$\lambda_{\tau_1}$}};
}
\node[anchor=center] at (1.375,0.375){$=$};
\node[anchor=north] at (2,-0.525){\footnotesize{$1$}};
\node[anchor=north] at (2.75,-0.525){\footnotesize{$2$}};
\node[anchor=west] at (3.25,0.75){\footnotesize{$3$}};
\draw[black,line width=0.4mm] (1.5,0.75) -- (3.25,0.75);
\draw[fill=red!10!green!50!blue!20, opacity=1, rounded corners = 2,thick] (1.875,-0.25) rectangle ++(1,0.5);
\node[anchor=center] at (2.375,0){$\tau$};
\draw[green!20!blue, line width=1mm, opacity=0.5] (2,1.25) -- (2,1);
\draw[green!20!blue, line width=1mm, opacity=0.5] (2,0.5) -- (2,0.25);
\draw[red!80!blue, line width=0.5mm, opacity=0.5] (2.75,0.5) -- (2.75,0.25);
\draw[red!80!blue, line width=0.5mm, opacity=0.5] (2.75,1.25) -- (2.75,1);
\draw[red!80!blue, line width=0.5mm, opacity=0.5] (2,-0.25) -- (2,-0.5);
\draw[green!20!blue, line width=1mm, opacity=0.5] (2.75,-0.25) -- (2.75,-0.5);

\foreach \x in {2.5}
{
\draw[fill=white, opacity=1, rounded corners = 0.5,thick] (\x+0.25,0.75-0.3) -- (\x+0.25+0.3,0.75) -- (\x+0.25,0.75+0.3) -- (\x+0.25-0.3,0.75) -- (\x+0.25,0.75-0.3);
\draw[fill=red!80!blue, opacity=0.25, rounded corners = 0.5,thick] (\x+0.25,0.75-0.3) -- (\x+0.25+0.3,0.75) -- (\x+0.25,0.75+0.3) -- (\x+0.25-0.3,0.75) -- (\x+0.25,0.75-0.3);
\node[anchor=center] at (\x+0.25,0.75){\footnotesize{$\lambda_{\tau_1}$}};

\draw[fill=white, opacity=1, rounded corners = 0.5,thick] (\x-0.5,0.75-0.3) -- (\x-0.5+0.3,0.75) -- (\x-0.5,0.75+0.3) -- (\x-0.5-0.3,0.75) -- 
(\x-0.5,0.75-0.3);
\draw[fill=green!20!blue, opacity=0.25, rounded corners = 0.5,thick] (\x-0.5,0.75-0.3) -- (\x-0.5+0.3,0.75) -- (\x-0.5,0.75+0.3) -- (\x-0.5-0.3,0.75) -- (\x-0.5,0.75-0.3);
\node[anchor=center] at (\x-0.5,0.75){\footnotesize{$\lambda_{\tau_2}$}};
}
\end{tikzpicture}
\label{eq:ybe-diagram}
\end{equation}
will prove to be particularly useful. Here, the propagator $U$ is in green and labeled by the gate parameter $\tau$, the black horizontal line represents the auxiliary qubit, now denoted by $3$, and the colour of the $R$ matrices is related to the shift in their spectral parameter (${\color{red!80!blue}\tau_1}$ in red and ${\color{green!20!blue}\tau_2}$ in blue). The order of multiplication is from bottom to top, and the vertical lines represent qubits denoted by indices $1$ and $2$ in Eq.~\eqref{eq:gate-RLL}. Observe that the colour and the width of the vertical lines are also associated with the shifts in the spectral parameters: a thin red line stands for the shift ${\color{red!80!blue}\tau_1}$, and a thicker blue-one stands for ${\color{green!20!blue}\tau_2}$. As the $R$ matrices are carried over a gate, the shifts in their respective spectral parameters get exchanged.

As described below, we can use the diagrammatic representation~\eqref{eq:ybe-diagram} of the YBE to identify a family of commuting transfer matrices $\mathds{T}(\lambda)$ generating a given integrable circuit. As a warm-up exercise we first do this for the integrable first-order Trotterization of the Heisenberg model, worked out in Ref.~\cite{vanicat2018}. Its discrete-time propagator is
\begin{equation}
    \mathds{U}=\prod_{i=1}^{\frac{L}{2}} U_{2i,2i+1}\prod_{j=1}^{\frac{L}{2}} U_{2j-1,2j},
    \label{eq:XXX-trotter}
\end{equation}
where periodic boundary conditions are imposed by requiring that the indices $1$ and $L+1$ denote the same site. The name ``Trotterization'' is justified due to the Trotter-Suzuki limit: fixing a finite time $t=n \tau$, where $n\in\mathbb{N}$, the continuous-time evolution with the Heisenberg Hamiltonian $H=\sum_{j=1}^L P_{j,j+1}$ is reproduced as
\begin{equation}
    e^{-\ii t H}=\lim_{\tau\to 0}\mathds{U}^{t/\tau},
\end{equation}
where the limit implies $n\to \infty$, such that $t$ remains fixed.

Using Eq.~\eqref{eq:ybe-diagram} in the diagrammatic representation of the Trotterized Heisenberg model~\eqref{eq:XXX-trotter}, we can now recognize the transfer matrix $\mathds{T}(\lambda)$ that commutes with the circuit:\footnote{For simplicity we have fixed the number of lattice sites to $L=10$, but the diagrammatic argument extends to any even positive integer length $L$.}
\begin{equation}
\begin{tikzpicture}[baseline=(current  bounding  box.center),scale=1]
\foreach \x in {0,...,5}
    {
    \draw[fill=red!10!green!50!blue!20, opacity=1, rounded corners = 2,thick] (-0.125+\x,0) rectangle ++(0.75,0.5);
    \draw[fill=red!10!green!50!blue!20, opacity=1, rounded corners = 2,thick] (-0.625+\x,-0.75) rectangle ++(0.75,0.5);
    }
\fill[white] (-0.75,-1) rectangle ++(1,2);
\fill[white] (5.25,-1) rectangle ++(1,2);
\foreach \x in {0,...,4}
{
\draw[red!80!blue, line width=0.5mm, opacity=0.5] (0.5+\x,-1) -- (0.5+\x,-0.75);
\draw[green!20!blue, line width=1mm, opacity=0.5] (1+\x,-1) -- (1+\x,-0.75);
\draw[green!20!blue, line width=1mm, opacity=0.5] (0.5+\x,-0.25) -- (0.5+\x,0);
\draw[red!80!blue, line width=0.5mm, opacity=0.5] (1+\x,-0.25) -- (1+\x,0);
\draw[red!80!blue, line width=0.5mm, opacity=0.5] (0.5+\x,0.5) -- (0.5+\x,0.75);
\draw[green!20!blue, line width=1mm, opacity=0.5] (1+\x,0.5) -- (1+\x,0.75);
}

\draw[black,line width=0.4mm] (0.125,-1.125) to[out=180,in=90] (0,-1.125-0.125);
\draw[black,line width=0.4mm]  (0,-1.125-0.125) to[out=-90,in=180] (0.125,-1.125-0.25);
\draw[black,line width=0.4mm] (0.125,-1.125) -- (5.375,-1.125);
\draw[black,line width=0.4mm] (0.125,-1.125-0.25) -- (5.375,-1.125-0.25);
\draw[black,line width=0.4mm] (5.375,-1.125) to[out=0,in=90] (5.5,-1.125-0.125);
\draw[black,line width=0.4mm] (5.375,-1.125-0.25) to[out=0,in=-90] (5.5,-1.125-0.125);

\foreach \x in {1,...,5}
{
\draw[fill=white, opacity=1, rounded corners = 0.5,thick] (\x,-1.125-0.2) -- (\x+0.2,-1.125) -- (\x,-1.125+0.2) -- (\x-0.2,-1.125) -- (\x,-1.125-0.2);
\draw[fill=green!20!blue, opacity=0.25, rounded corners = 0.5,thick] (\x,-1.125-0.2) -- (\x+0.2,-1.125) -- (\x,-1.125+0.2) -- (\x-0.2,-1.125) -- (\x,-1.125-0.2);

\draw[fill=white, opacity=1, rounded corners = 0.5,thick] (\x-0.5,-1.125-0.2) -- (\x-0.5+0.2,-1.125) -- (\x-0.5,-1.125+0.2) -- (\x-0.5-0.2,-1.125) -- 
(\x-0.5,-1.125-0.2);
\draw[fill=red!80!blue, opacity=0.25, rounded corners = 0.5,thick] (\x-0.5,-1.125-0.2) -- (\x-0.5+0.2,-1.125) -- (\x-0.5,-1.125+0.2) -- (\x-0.5-0.2,-1.125) -- (\x-0.5,-1.125-0.2);
}

\draw[-{Stealth[scale=0.75]}, black, line width=0.5mm] (-0.125,-1.5) -- (-0.125,-0.5);
\draw[-{Stealth[scale=0.75]}, black, line width=0.5mm] (-0.125,-1.5) -- (1-0.125,-1.5);
\node[anchor=east] at (0-0.125,-0.5){$t$};
\node[anchor=north] at (1-0.125,-1.5){$x$};
\foreach \x in {1,...,10}
{
\node[anchor=south] at (0.5*\x,0.75){\footnotesize{\x}};
}
\end{tikzpicture}
\label{eq:XXX-trotter-diagram}
\end{equation}
Specifically, by applying the YBE diagram~\eqref{eq:ybe-diagram}, we can transfer the sequence of red and blue diamonds connected with a curved black line from the bottom of the circuit to the top, without changing their ordering---the ingoing and outgoing sequences of thin red and thick blue vertical lines are the same. That sequence of diamonds is the transfer matrix, the closed black curve representing the partial trace over the auxiliary qubit. From the colours of the diamonds [see Eq.~\eqref{eq:ybe-diagram}] or, equivalently, from the widths of the ingoing (or outgoing) vertical lines, we can identify the sequence of inhomogeneities that enter the transfer matrix~\eqref{eq:transfer-matrix} on $L$ sites. It reads
\begin{equation}
    \bs{\nu}=({\color{red!80!blue}\tau_1},{\color{green!20!blue}\tau_2},{\color{red!80!blue}\tau_1},{\color{green!20!blue}\tau_2},{\color{red!80!blue}\tau_1},{\color{green!20!blue}\tau_2},\ldots,
    {\color{red!80!blue}\tau_1},{\color{green!20!blue}\tau_2},{\color{red!80!blue}\tau_1},{\color{green!20!blue}\tau_2}).
    \label{eq:staggered-T}
\end{equation} 

The Floquet propagator~\eqref{eq:XXX-trotter} of the Trotterized Heisenberg model, as well as the two-site periodic lattice shift towards the west (left), are then reproduced as
\begin{equation}
    \mathds{U}=[\mathds{T}(\tau_1)]^{-1}\mathds{T}(\tau_2),\qquad \mathds{W}^2=\mathds{T}(\tau_1)\mathds{T}(\tau_2),
    \label{eq:U-and-W}
\end{equation}
respectively. Here, the one-site periodic lattice shift towards the west is 
\begin{equation}
    \mathds{W}=P_{1,L}P_{2,L}\cdots P_{L-1,L}=P_{1,2}P_{1,3}\cdots P_{1,L}.
    \label{eq:west-shift}
\end{equation}
It acts on the product states as $\mathds{W}\ket{s_1,s_2,\ldots s_{L-1},s_L}=\ket{s_2,s_3,\cdots,s_L,s_1}$, where $s_j\in\{\frac{1}{2},-\frac{1}{2}\}$, and it shifts the operators as, e.g., $\mathds{W}U_{j,j+1}=U_{j-1,j}\mathds{W}$.

To summarize, we have used diagrams to identify the transfer matrix $\mathds{T}(\lambda)$ that commutes with the propagator $\mathds{U}$. The transfer matrix consists of $R$ matrices, whose spectral parameters are shifted either by $\tau_1$, or by $\tau_2$ (the so-called inhomogeneities). We indicate the shift of the spectral parameter in each $R$ matrix by making the vertical line that represents a qubit either thin and red (${\color{red!80!blue}\tau_1}$), or thick and blue (${\color{green!20!blue}\tau_2}$). According to Eq.~\eqref{eq:ybe-diagram}, the gate exchanges the colours and thicknesses. If a sequence of colours and thicknesses is preserved after applying the full propagator $\mathds{U}$, we can associate a transfer matrix satisfying $[\mathds{T}(\lambda),\mathds{U}]=0$ with it, and hence assume that the circuit is integrable. 

\section{Generalization to other geometries}
\label{sec:gen_geometries}

The above reasoning can be used to identify other integrable arrangements of quantum gates. To this end, consider a single qubit in the bulk of an integrable circuit, say, the one in the diagram~\eqref{eq:XXX-trotter-diagram}. If acted upon by the right leg of a quantum gate, the vertical line representing it changes from thick blue to thin red. Conversely, it changes from thin red to thick blue, if it is subject to the action of the left leg of a quantum gate:
\begin{equation}
\begin{tikzpicture}[baseline=(current  bounding  box.center),scale=1.25]
\draw[fill=red!10!green!50!blue!20, opacity=1, rounded corners = 2,thick] (-0.125,0.5) rectangle ++(1,0.5);
\draw[green!20!blue, line width=1mm, opacity=0.5] (0,1.25) -- (0,1);
\draw[green!20!blue, line width=1mm, opacity=0.5] (0.75,0.5) -- (0.75,0.25);
\draw[red!80!blue, line width=0.5mm, opacity=0.5] (0.75,1.25) -- (0.75,1);
\draw[red!80!blue, line width=0.5mm, opacity=0.5] (0,0.5) -- (0,0.25);
\fill[white] (-0.5,0) rectangle ++(1,1.5);
\node[anchor=center] at (0.125,0.75){$\ldots$};
\node[anchor=center] at (1.25,0.75){$\ldots$};
\draw[fill=red!10!green!50!blue!20, opacity=1, rounded corners = 2,thick] (3-0.125,0.5) rectangle ++(1,0.5);
\draw[green!20!blue, line width=1mm, opacity=0.5] (3,1.25) -- (3,1);
\draw[green!20!blue, line width=1mm, opacity=0.5] (3.75,0.5) -- (3.75,0.25);
\draw[red!80!blue, line width=0.5mm, opacity=0.5] (3.75,1.25) -- (3.75,1);
\draw[red!80!blue, line width=0.5mm, opacity=0.5] (3,0.5) -- (3,0.25);
\fill[white] (3.25,0) rectangle ++(1,1.5);
\node[anchor=center] at (2.5,0.75){$\ldots$};
\node[anchor=center] at (3.625,0.75){$\ldots$};
\draw[-{Stealth[scale=0.75]}, black, line width=0.5mm] (-0.375,0.25) -- (-0.375,1.25);
\node[anchor=east] at (-0.375,0.75) {$t$};
\end{tikzpicture}
\label{eq:qubit-color-change}
\end{equation}

Suppose now, that the time-period propagator $\mathds{U}$ commutes with a transfer matrix $\mathds{T}(\lambda)$. The circuit then preserves the sequence of line colours and thicknesses: the latter corresponds to the sequence $\bs{\nu}$ of inhomogeneities in the transfer matrix. The thickness and colour of each qubit's vertical line in such a circuit should change an even number of times. In the simplest case it should change twice in a period: {\color{red!80!blue} thin red} $\to$ {\color{green!20!blue}thick blue} $\to$ {\color{red!80!blue}thin red}, or {\color{green!20!blue}thick blue} $\to$ {\color{red!80!blue}thin red} $\to$ {\color{green!20!blue}thick blue}. Considering this simplest scenario, in one period $\mathds{U}$, each qubit should be acted upon once by the left leg of a gate, and once by its right leg---cf. Eq.~\eqref{eq:qubit-color-change}. This certainly occurs in the same-gate circuits in which each neighbouring pair of qubits is acted upon by the gate exactly once in a period. This is because each qubit forms two distinct pairs---one with the qubit on its left-hand side, the other with the qubit on its right-hand side. Therefore, we have the following conjecture:

\begin{theorem} Let $U$ be a gate of the form~\eqref{eq:unitary-gate}, satisfying Eq.~\eqref{eq:gate-RLL}. A time-periodic circuit, in which $U$ is applied to each pair of neighbouring qubits exactly once per period, is integrable.
\end{theorem}

The one-period propagator of such a circuit is reproduced as $\mathds{U}=[\mathds{T}(\tau_1)]^{-1}\mathds{T}(\tau_2)$, where the inhomogeneities $\bs{\nu}$ in the transfer matrix~\eqref{eq:transfer-matrix} are identified from the conserved sequence of ingoing (or outgoing) thin red ($\nu_j=\tau_1$) and thick blue ($\nu_j=\tau_2$) lines. Whether $\mathds{U}$ is really reproduced from the transfer matrix in such a way has to be checked separately for each case. This is why the above claim on integrability remains a conjecture. In what follows we will first consider integrable circuits on $L=10$ qubits, classified in Ref.~\cite{duhClassificationSamegateQuantum2024} (see Fig. 4 therein and Fig. \ref{fig:circuits} below).

\begin{figure}[ht!]
    \centering
    \hspace{3.5em}
    \includegraphics[width=0.8\textwidth]{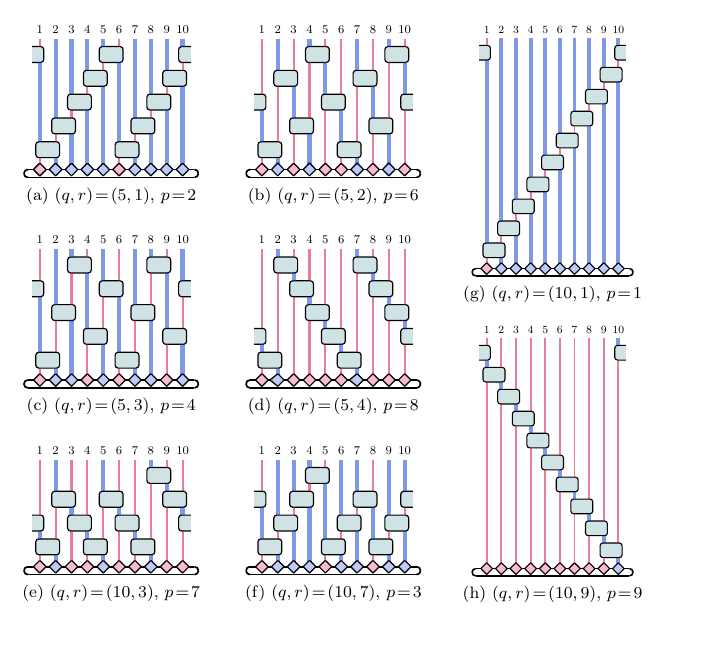}
    \vspace{-3em}
    \caption{Representative integrable circuits on $L=10$ qubits: any circuit on $L=10$ qubits, whose gate acts on every pair of neighbouring qubits exactly once per period, has the same spectrum as either one of these circuits, or as a brickwork circuit in Eq.~\eqref{eq:XXX-trotter-diagram}. The coloured lines of different widths indicate what occurs to $R$ matrices that constitute the transfer matrix (a sequence of diamonds at the bottom of each panel, connected by a closed loop), once they are exchanged with a gate according to YBE~\eqref{eq:ybe-diagram}. The transfer matrix is preserved if the output sequence of colours and widths (top of each panel) matches the input-one (bottom of each panel). The sequence of line widths and colours is equivalent to the sequence of inhomogeneities ${\color{red!80!blue}\tau_1}$ (thick, red), ${\color{green!20!blue}\tau_2}$ (thick, blue), i.e., the shifts in the spectral parameters of the $R$ matrices.}
    \label{fig:circuits}
\end{figure}

\subsection{Classification of integrable circuits for L {\rm = 10}}
\label{sec:classification-L10}

There are $L-1$ SECs for any fixed system size $L$~\cite{duhClassificationSamegateQuantum2024}. The first class contains the brickwork circuit, described in Section~\ref{sec:brickwork}. The eight representatives of the remaining SECs on $L=10$ qubits are depicted in Fig.~\ref{fig:circuits}. Below we will describe the four given in Figs.~\hyperref[fig:circuits]{1(a)}, \hyperref[fig:circuits]{1(b)}, \hyperref[fig:circuits]{1(f)}, and \hyperref[fig:circuits]{1(h)}. The rest can be mapped into these examples or worked-out similarly. For example, the transfer matrices of circuits in Fig.~\hyperref[fig:circuits]{1(a)} and \hyperref[fig:circuits]{1(d)} are related by a lattice shift, followed by an exchange of the parameters $\tau_1$ and $\tau_2$ (i.e., exchange of blue and red diamonds or, equivalently, thin red and thick blue lines)\footnote{Note that, due to the exchange of parameters $\tau_1$ and $\tau_2$, the spectra are not the same. These two circuits are therefore not in the same SEC.}. Other circuits can be paired in a similar fashion.
\begin{enumerate}
    \item[\textbf{(a)}] \emph{Class $(q,r)=(5,1)$ (equiv. $p=2$)}: The conserved sequence of thin red and thick blue lines (or red and blue diamonds) in Fig.~\hyperref[fig:circuits]{1(a)} suggests the following sequence of inhomogeneities in the transfer matrix~\eqref{eq:transfer-matrix}:
    $\bs{\nu}=({\color{red!80!blue}\tau_1},{\color{green!20!blue}\tau_2},{\color{green!20!blue}\tau_2},{\color{green!20!blue}\tau_2},{\color{green!20!blue}\tau_2},{\color{red!80!blue}\tau_1},{\color{green!20!blue}\tau_2},{\color{green!20!blue}\tau_2},{\color{green!20!blue}\tau_2},{\color{green!20!blue}\tau_2})$.
    Note that there are exactly $p=2$ inhomogeneities $\tau_1$. Using $U=P R(\tau_2\!-\!\tau_1)$, we obtain 
    $\mathds{T}(\tau_1)=\mathds{W}\,U_{2,3}^{-1}U_{3,4}^{-1}U_{4,5}^{-1}U_{5,6}^{-1}U_{7,8}^{-1}U_{8,9}^{-1}U_{9,10}^{-1}U_{10,1}^{-1}$ and $\mathds{T}(\tau_2)=\mathds{W}\,U_{1,2}^{}U_{6,7}^{}$,
    where $\mathds{W}$ is a periodic lattice shift towards the west, defined in Eq.~\eqref{eq:west-shift}. Combining the two transfer matrices yields
    \begin{equation}
    [\mathds{T}(\tau_1)]^{-1}\mathds{T}(\tau_2)=[U_{10,1}U_{5,6}][U_{9,10}U_{4,5}][U_{8,9}U_{3,4}][U_{7,8}U_{2,3}][U_{6,7}U_{1,2}],
    \end{equation}
    which is exactly the propagator $\mathds{U}$ for one period in time, depicted in Fig.~\hyperref[fig:circuits]{1(a)}. Moreover, using $\mathds{W}U_{j,j+1}=U_{j-1,j}\mathds{W}$, we find
    \begin{equation}
        \mathds{T}({\color{red!80!blue}\tau_1})[\mathds{T}({\color{green!20!blue}\tau_1})]^4=\mathds{W}^5.  
    \end{equation}
    This implies that the circuit is invariant under periodic translations for $q=5$ sites, since transfer matrices commute with the propagator $\mathds{U}$.

    \item[\textbf{(b)}] \emph{Class $(q,r)=(5,2)$ (equiv. $p=6$)}: The transfer matrix in Fig.~\hyperref[fig:circuits]{1(b)} now has $p=6$ inhomogeneities $\tau_1$:
    $\bs{\nu}=({\color{red!80!blue}\tau_1},{\color{green!20!blue}\tau_2},{\color{red!80!blue}\tau_1},{\color{green!20!blue}\tau_2},{\color{red!80!blue}\tau_1},{\color{red!80!blue}\tau_1},{\color{green!20!blue}\tau_2},{\color{red!80!blue}\tau_1},{\color{green!20!blue}\tau_2},{\color{red!80!blue}\tau_1})$.
    From
    $\mathds{T}(\tau_1)=\mathds{W}\,U_{2,3}^{-1}U_{4,5}^{-1}U_{7,8}^{-1}U_{9,10}^{-1}$, and $\mathds{T}(\tau_2)=\mathds{W}\,U_{10,1}^{}U_{1,2}^{}U_{3,4}^{}U_{5,6}^{}U_{6,7}^{}U_{8,9}^{}$,
    we obtain the propagator
    \begin{equation}
        [\mathds{T}(\tau_1)]^{-1}\mathds{T}(\tau_2)=[U_{9,10}^{}U_{4,5}^{}][U_{7,8}^{}U_{2,3}^{}][U_{10,1}^{}U_{5,6}^{}][U_{3,4}^{}U_{8,9}^{}][U_{1,2}^{}U_{6,7}^{}],
    \end{equation}
    depicted in Fig.~\hyperref[fig:circuits]{1(b)}, as well as the lattice shift for $q=5$ sites
    \begin{equation}
        \mathds{T}({\color{red!80!blue}\tau_1})\mathds{T}({\color{green!20!blue}\tau_2})\mathds{T}({\color{red!80!blue}\tau_1})\mathds{T}({\color{green!20!blue}\tau_2})\mathds{T}({\color{red!80!blue}\tau_1})=\mathds{W}^5.
    \end{equation}
    The latter is again a product of $q=5$ transfer matrices, evaluated at the inhomogeneities in the spatial period $({\color{red!80!blue}\tau_1},{\color{green!20!blue}\tau_2},{\color{red!80!blue}\tau_1},{\color{green!20!blue}\tau_2},{\color{red!80!blue}\tau_1})$.

    \item[\textbf{(f)}] \emph{Class $(q,r)=(10,7)$ (equiv. $p=3$)}: There are $p=3$ inhomogeneities $\tau_1$ in the transfer matrix identified from Fig.~\hyperref[fig:circuits]{1(f)}:
    $\bs{\nu}=({\color{red!80!blue}\tau_1},{\color{green!20!blue}\tau_2},{\color{green!20!blue}\tau_2},{\color{green!20!blue}\tau_2},{\color{red!80!blue}\tau_1},{\color{green!20!blue}\tau_2},{\color{green!20!blue}\tau_2},{\color{red!80!blue}\tau_1},{\color{green!20!blue}\tau_2},{\color{green!20!blue}\tau_2})$.
    We have
    $\mathds{T}(\tau_1)=\mathds{W}\,U_{2,3}^{-1}U_{3,4}^{-1}U_{4,5}^{-1}U_{6,7}^{-1}U_{7,8}^{-1}U_{9,10}^{-1}U_{10,1}^{-1}$, and $\mathds{T}(\tau_2)=\mathds{W}\,U_{1,2}^{}U_{5,6}^{}U_{8,9}^{}$,
    yielding the propagator $\mathds{U}$ depicted in Fig.~\hyperref[fig:circuits]{1(f)}:
    \begin{equation}
        [\mathds{T}(\tau_1)]^{-1}\mathds{T}(\tau_2)=U_{4,5}^{}[U_{10,1}^{}U_{7,8}^{}U_{3,4}^{}][U_{9,10}^{}U_{6,7}^{}U_{2,3}^{}][U_{8,9}^{}U_{5,6}^{}U_{1,2}^{}],
    \end{equation}
    In this example, the sequence of inhomogeneities does not have a smaller periodic subsequence. Multiplying the transfer matrices evaluated at all inhomogeneities, we obtain the lattice shift for $q=10$ sites:
    \begin{equation}
        [\mathds{T}({\color{red!80!blue}\tau_1})]^3[\mathds{T}({\color{green!20!blue}\tau_2})]^7=\mathds{W}^{10}.
    \end{equation}
    The only periodicity is the trivial one, due to the periodic boundary conditions.

    \item[\textbf{(h)}] \emph{Class $(q,r)=(10,9)$ (equiv. $p=9$)}: From  Fig.~\hyperref[fig:circuits]{1(h)} we identify the sequence of inhomogeneities as
    $\bs{\nu}=({\color{red!80!blue}\tau_1},{\color{red!80!blue}\tau_1},{\color{red!80!blue}\tau_1},{\color{red!80!blue}\tau_1},{\color{red!80!blue}\tau_1},{\color{red!80!blue}\tau_1},{\color{red!80!blue}\tau_1},{\color{red!80!blue}\tau_1},{\color{red!80!blue}\tau_1},{\color{green!20!blue}\tau_2})$.
    The transfer matrix yields
    $\mathds{T}(\tau_1)=\mathds{W}\,U_{10,1}^{-1}$ and
    $\mathds{T}(\tau_2)=\mathds{W}\,U_{1,2}^{}U_{2,3}^{}U_{3,4}^{}U_{4,5}^{}U_{5,6}^{}U_{6,7}^{}U_{7,8}^{}U_{8,9}^{}U_{9,10}^{}$.
    The staircase propagator in Fig.~\hyperref[fig:circuits]{1(h)} is reproduced as
    \begin{equation}
        [\mathds{T}(\tau_1)]^{-1}\mathds{T}(\tau_2)=U_{10,1}^{}U_{1,2}^{}U_{2,3}^{}U_{3,4}^{}U_{4,5}^{}U_{5,6}^{}U_{6,7}^{}U_{7,8}^{}U_{8,9}^{}U_{9,10}^{},
    \end{equation}
    and again there is only the trivial periodicity due to the periodic boundary conditions:
    \begin{equation}
        [\mathds{T}(\tau_1)]^9\mathds{T}(\tau_2)=\mathds{W}^{10}.
    \end{equation}
\end{enumerate}

Observe that, in each case, the SEC classification number $p$ is the same as the number of inhomogeneities $\tau_1$ in the transfer matrix. The remaining $L-p$ inhomogeneities are $\tau_2$. As we will see later on, transfer matrices with the same set of inhomogeneities (i.e., irrespective of their ordering in the sequence $\nu$) share the same set of Bethe equations and thus have the same spectrum---see subsection~\ref{sec:bethe-eqs}. This is moreover consistent with the observation in Ref.~\cite{duhClassificationSamegateQuantum2024}, that the circuits sharing the same spectrum of quasienergies are unitarily equivalent to the circuit 
\begin{equation}
    \mathds{U}=\left(\prod_{p+1\le m\le L}^{\leftarrow} U_{m,m+1}\right)\left(\prod_{1\le \ell \le p}^{\rightarrow} U_{\ell,\ell+1} \right),
    \label{eq:double-staircase}
\end{equation}
composed of two staircases with opposite chirality, with respective lengths $p$ and $L-p$. In our case, an integrable circuit of this sort originates in the transfer matrix~\eqref{eq:transfer-matrix}, in which we set the inhomogeneities to be $\nu_{j}=\tau_1$ for $1\le j\le p$ and $\nu_{j}=\tau_2$ for $p<j\le L$. Other circuits in the same SEC as~\eqref{eq:double-staircase} can be obtained by applying a suitable sequence of permutations that redistribute the inhomogeneities inside the transfer matrix.

\subsection{Integrable circuits with differing layers}
\label{sec:different-layers}

We now go beyond the same-gate circuits by considering transfer matrices with periodic patterns of multiple inhomogeneities. Consider, for example, a circuit with space periodicity $q=3$ (for simplicity we now also assume $L\in q\mathbb{N}$). Take
\begin{equation}
    \bs{\nu}=({\color{green!50!blue}\tau_3},{\color{green!20!blue}\tau_2},{\color{red!80!blue}\tau_1},{\color{green!50!blue}\tau_3},{\color{green!20!blue}\tau_2},{\color{red!80!blue}\tau_1},\ldots,{\color{green!50!blue}\tau_3},{\color{green!20!blue}\tau_2},{\color{red!80!blue}\tau_1})
    \label{eq:3-inhomogeneities}
\end{equation}
for the sequence of inhomogeneities in the transfer matrix~\eqref{eq:transfer-matrix}. Invoking the YBE diagram~\eqref{eq:ybe-diagram} again, we observe that such a transfer matrix commutes with a circuit, in which different layers are formed from different gates. Specifically, on $L=12$ qubits such a circuit looks as follows:
\begin{equation}
\begin{tikzpicture}[baseline=(current  bounding  box.center),scale=1]
\foreach \x in {-1,...,2}
    {
    \draw[fill=red!10!green!50!blue!20, opacity=1, rounded corners = 2,thick] (0.375+1.5*\x,1.5) rectangle ++(0.75,0.5);
    \draw[pattern=crosshatch, pattern color=blue!80!green!80, opacity=1, rounded corners = 2,thick] (-0.625+1.5*\x,0.75) rectangle ++(0.75,0.5);
    \draw[pattern=north west lines, pattern color=red!70!blue!80, opacity=1, rounded corners = 2,thick] (0.375+1.5*\x,0) rectangle ++(0.75,0.5);
    \draw[pattern=crosshatch, pattern color=blue!80!green!80, opacity=1, rounded corners = 2,thick] (-0.125+1.5*\x,-0.75) rectangle ++(0.75,0.5);
    }
\draw[pattern=crosshatch, pattern color=blue!80!green!80, opacity=1, rounded corners = 2,thick] (-0.625+1.5*3,0.75) rectangle ++(0.75,0.5);

\fill[white] (-2.75,-1) rectangle ++(1,3);
\fill[white] (4.25,-1) rectangle ++(1,3);

\foreach \x in {-1,...,2}
{
\draw[red!80!blue, line width=0.25mm, opacity=0.5] (1+1.5*\x,-1) -- (1+1.5*\x,0);
\draw[red!80!blue, line width=0.25mm, opacity=0.5] (0.5+1.5*\x,0.5) -- (0.5+1.5*\x,1.5);
\draw[red!80!blue, line width=0.25mm, opacity=0.5] (1+1.5*\x,2) -- (1+1.5*\x,2.25);
\draw[green!20!blue, line width=0.7mm, opacity=0.5] (0.5+1.5*\x,-1) -- (0.5+1.5*\x,-0.75);
\draw[green!20!blue, line width=0.7mm, opacity=0.5] (1.5*\x,-0.25) -- (1.5*\x,0.75);
\draw[green!20!blue, line width=0.7mm, opacity=0.5] (1+1.5*\x,1.25) -- (1+1.5*\x,1.5);
\draw[green!20!blue, line width=0.7mm, opacity=0.5] (0.5+1.5*\x,2) -- (0.5+1.5*\x,2.25);
%
\draw[green!50!black, line width=1.15mm, opacity=0.75] (1.5*\x,-1) -- (1.5*\x,-0.75);
\draw[green!50!black, line width=1.15mm, opacity=0.75] (1.5*\x,1.25) -- (1.5*\x,2.25);
\draw[green!50!black, line width=1.15mm, opacity=0.75] (0.5+1.5*\x,-0.25) -- (0.5+1.5*\x,0);
\draw[green!50!black, line width=1.15mm, opacity=0.75] (1+1.5*\x,0.5) -- (1+1.5*\x,0.75);
}

\draw[black,line width=0.4mm] (-2+0.125,-1.125) to[out=180,in=90] (-2,-1.125-0.125);
\draw[black,line width=0.4mm]  (-2,-1.125-0.125) to[out=-90,in=180] (-2+0.125,-1.125-0.25);
\draw[black,line width=0.4mm] (-2+0.125,-1.125) -- (-1+5.375,-1.125);
\draw[black,line width=0.4mm] (-2+0.125,-1.125-0.25) -- (-1+5.375,-1.125-0.25);
\draw[black,line width=0.4mm] (-1+5.375,-1.125) to[out=0,in=90] (-1+5.5,-1.125-0.125);
\draw[black,line width=0.4mm] (-1+5.375,-1.125-0.25) to[out=0,in=-90] (-1+5.5,-1.125-0.125);

\foreach \x in {0,...,3}
{
\draw[fill=white, opacity=1, rounded corners = 0.5,thick] (1.5*\x-0.5,-1.125-0.2) -- (1.5*\x-0.5+0.2,-1.125) -- (1.5*\x-0.5,-1.125+0.2) -- (1.5*\x-0.5-0.2,-1.125) -- (1.5*\x-0.5,-1.125-0.2);
\draw[fill=red!80!blue, opacity=0.25, rounded corners = 0.5,thick] (1.5*\x-0.5,-1.125-0.2) -- (1.5*\x-0.5+0.2,-1.125) -- (1.5*\x-0.5,-1.125+0.2) -- (1.5*\x-0.5-0.2,-1.125) -- (1.5*\x-0.5,-1.125-0.2);
\draw[fill=white, opacity=1, rounded corners = 0.5,thick] (1.5*\x-1,-1.125-0.2) -- (1.5*\x-1+0.2,-1.125) -- (1.5*\x-1,-1.125+0.2) -- (1.5*\x-1-0.2,-1.125) -- (1.5*\x-1,-1.125-0.2);
\draw[fill=green!20!blue, opacity=0.25, rounded corners = 0.5,thick] (1.5*\x-1,-1.125-0.2) -- (1.5*\x-1+0.2,-1.125) -- (1.5*\x-1,-1.125+0.2) -- (1.5*\x-1-0.2,-1.125) -- (1.5*\x-1,-1.125-0.2);
\draw[fill=green!40!blue!20!white, opacity=1, rounded corners = 0.5,thick] (1.5*\x-1.5,-1.125-0.2) -- (1.5*\x-1.5+0.2,-1.125) -- (1.5*\x-1.5,-1.125+0.2) -- (1.5*\x-1.5-0.2,-1.125) -- (1.5*\x-1.5,-1.125-0.2);
\draw[pattern=crosshatch, pattern color=green!50!blue, opacity=1, rounded corners = 0.5,thick] (1.5*\x-1.5,-1.125-0.2) -- (1.5*\x-1.5+0.2,-1.125) -- (1.5*\x-1.5,-1.125+0.2) -- (1.5*\x-1.5-0.2,-1.125) -- (1.5*\x-1.5,-1.125-0.2);
}

\draw[-{Stealth[scale=0.75]}, black, line width=0.5mm] (-2.125,-1.5) -- (-2.125,-0.5);
\draw[-{Stealth[scale=0.75]}, black, line width=0.5mm] (-2.125,-1.5) -- (1-2.125,-1.5);
\node[anchor=east] at (-2.125,-0.5){$t$};
\node[anchor=north] at (1-2.125,-1.5){$x$};
\node[anchor=east] at (-1.5,1.75){\color{green!50!blue}$\tau_3$};
\node[anchor=east] at (-1.5,0.25){\color{green!20!blue}$\tau_2$};
\node[anchor=west] at (4,-0.5){\color{red!80!blue}$\tau_1$};
\draw[black,thick] (6.325,0.875) -- (6.325,1.5);
\draw[black,thick] (6.675,0.875) -- (6.675,1.5);
\draw[fill=red!10!green!50!blue!20, opacity=1, rounded corners = 2,thick] (6.25,1) rectangle ++(0.5,0.375);
\draw[black,thick] (6.325,0.125) -- (6.325,0.75);
\draw[black,thick] (6.675,0.125) -- (6.675,0.75);
\draw[fill=white, opacity=1, rounded corners = 2,thick] (6.25,0.25) rectangle ++(0.5,0.375);
\draw[pattern=crosshatch, pattern color=blue!80!green!80, opacity=1, rounded corners = 2,thick] (6.25,0.25) rectangle ++(0.5,0.375);
\draw[black,thick] (6.325,-0.625) -- (6.325,0);
\draw[black,thick] (6.675,-0.625) -- (6.675,0);
\draw[fill=white, opacity=1, rounded corners = 2,thick] (6.25,-0.5) rectangle ++(0.5,0.375);
\draw[pattern=north west lines, pattern color=red!70!blue!80, opacity=1, rounded corners = 2,thick] (6.25,-0.5) rectangle ++(0.5,0.375);
\node[anchor=west] at (7,1.25){$U^{(\tau_{21})}$};
\node[anchor=west] at (7,0.5){$U^{(\tau_{23})}$};
\node[anchor=west] at (7,-0.25){$U^{(\tau_{13})}$};
\node[anchor=west] at(6,1.875){Gates:};
\node[anchor=west] at(6,-1){$\tau_{ij}\equiv\tau_i-\tau_j$};
\end{tikzpicture}
\label{eq:different-gate-circuit}
\end{equation}
The width and the colour of the vertical lines, which represent qubits, again indicate the exchange of inhomogeneities according to the YBE~\eqref{eq:ybe-diagram}. The new inhomogeneity, ${\color{green!50!blue}\tau_3}$, is indicated by the thickest (green) line. We should stress that the gate parameters in Eq.~\eqref{eq:different-gate-circuit} are differences of the inhomogeneities, i.e.,
\begin{equation}
    U^{(\tau_{ij})}\equiv P R(\tau_i-\tau_j).
\end{equation}
In particular, this means that not every circuit with the above geometry is integrable---the parameters in the $R$ matrices which correspond to gates should be properly related. 

\begin{figure}[ht!]
    \centering
    \hspace{1em}
    \includegraphics[width=0.9\textwidth]{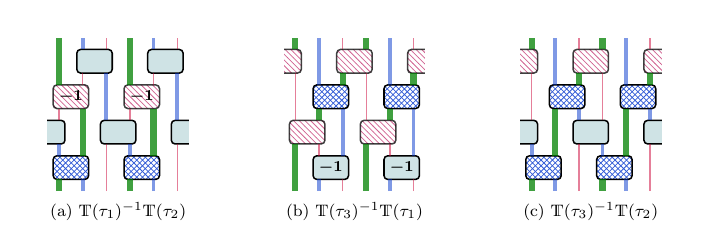}
    \vspace{-1em}
    \caption{Three circuits obtained from transfer matrices~\eqref{eq:three-transfer-matrices}. The gates are applied from bottom to top. Symbol $\bs{-1}$ signifies the inverse of a quantum gate. The gates are otherwise the same as in Eq.~\eqref{eq:different-gate-circuit}. The right-most circuit is equivalent to the one in Eq.~\eqref{eq:different-gate-circuit}.}
    \label{fig:three-circuits}
\end{figure}

To construct the circuit in Eq.~\eqref{eq:different-gate-circuit} using the transfer matrix with inhomogeneities~\eqref{eq:3-inhomogeneities}, we first observe
\begin{eqnarray}
    &\mathds{T}(\tau_1)=\mathds{W}\prod_{1\le j\le \frac{L}{3}}^{\rightarrow}U_{3j-2,3j-1}^{(\tau_{13})}U_{3j-1,3j}^{(\tau_{12})},\nonumber\\
    &\mathds{T}(\tau_2)=\mathds{W}\prod_{1\le j\le \frac{L}{3}}^{\rightarrow}U_{3j-3,3j-2}^{(\tau_{21})}U_{3j-2,3j-1}^{(\tau_{23})},\\
    &\mathds{T}(\tau_3)=\mathds{W}\prod_{1\le j\le \frac{L}{3}}^{\rightarrow}U_{3j-1,3j}^{(\tau_{32})}U_{3j,3j+1}^{(\tau_{31})}.\nonumber
    \label{eq:three-transfer-matrices}
\end{eqnarray}
These transfer matrices can be combined to yield the three-site lattice shift $\mathds{T}(\tau_1)\mathds{T}(\tau_2)\mathds{T}(\tau_3)=\mathds{W}^3$, as well as the three circuits, whose bulk is depicted in Fig.~\ref{fig:three-circuits}. For example, recalling property~\eqref{eq:unitarity} of the $R$ matrix, which implies $[U^{(\tau_{ij})}]^{-1}=U^{(\tau_{ji})}$, we have [cf. Fig.~\hyperref[fig:three-circuits]{2(a)}]
\begin{equation}
    [\mathds{T}(\tau_1)]^{-1}\mathds{T}(\tau_2)=\prod_{1\le j\le \frac{L}{3}}^{\leftarrow}U_{3j-1,3j}^{(\tau_{21})}[U_{3j-2,3j-1}^{(\tau_{13})}]^{-1}\prod_{1\le j\le \frac{L}{3}}^{\rightarrow}U_{3j-3,3j-2}^{(\tau_{21})}U_{3j-2,3j-1}^{(\tau_{23})}.
\end{equation}
As usual, the arrows denote the direction of the increasing index in the product. 

As a consequence of YBE, the bulk of the circuit in Eq.~\eqref{eq:different-gate-circuit}, is equivalent to the one in Fig.~\hyperref[fig:three-circuits]{2(c)}, i.e., it is obtained as $\mathds{U}=[\mathds{T}(\tau_3)]^{-1}\mathds{T}(\tau_2)$. In particular, the gates parametrized by the differences of inhomogeneities satisfy the ``braided'' Yang-Baxter equation
\begin{equation}
(\mathds{1}\otimes U^{(\tau_{13})})(U^{(\tau_{23})}\otimes\mathds{1})(\mathds{1}\otimes U^{(\tau_{21})})=(U^{(\tau_{21})}\otimes\mathds{1})(\mathds{1}\otimes U^{(\tau_{23})})(U^{(\tau_{13})}\otimes\mathds{1}).
\end{equation}
Using its diagrammatic representation
\begin{equation}
\begin{tikzpicture}[baseline=(current  bounding  box.center),scale=1]
\draw[fill=red!10!green!50!blue!20, opacity=1, rounded corners = 2,thick](2.5+0.375-1.5,1.5) rectangle ++(0.75,0.5);
\draw[pattern=crosshatch, pattern color=blue!80!green!80, opacity=1, rounded corners = 2,thick] (2.5-0.625,0.75) rectangle ++(0.75,0.5);
\draw[pattern=north west lines, pattern color=red!70!blue!80, opacity=1, rounded corners = 2,thick] (2.5+0.375-1.5,0) rectangle ++(0.75,0.5);    
\foreach \x in {-1}
{
\draw[red!80!blue, line width=0.25mm, opacity=0.5] (2.5+1+1.5*\x,-0.25) -- (2.5+1+1.5*\x,0);
\draw[red!80!blue, line width=0.25mm, opacity=0.5] (2.5+0.5+1.5*\x,0.5) -- (2.5+0.5+1.5*\x,1.5);
\draw[red!80!blue, line width=0.25mm, opacity=0.5] (2.5+1+1.5*\x,2) -- (2.5+1+1.5*\x,2.25);
\draw[green!20!blue, line width=0.7mm, opacity=0.5] (2.5+1.5+1.5*\x,-0.25) -- (2.5+1.5+1.5*\x,0.75);
\draw[green!20!blue, line width=0.7mm, opacity=0.5] (2.5+1+1.5*\x,1.25) -- (2.5+1+1.5*\x,1.5);
\draw[green!20!blue, line width=0.7mm, opacity=0.5] (2.5+0.5+1.5*\x,2) -- (2.5+0.5+1.5*\x,2.25);
\draw[green!50!black, line width=1.15mm, opacity=0.75] (2.5+1.5+1.5*\x,1.25) -- (2.5+1.5+1.5*\x,2.25);
\draw[green!50!black, line width=1.15mm, opacity=0.75] (2.5+0.5+1.5*\x,-0.25) -- (2.5+0.5+1.5*\x,0);
\draw[green!50!black, line width=1.15mm, opacity=0.75] (2.5+1+1.5*\x,0.5) -- (2.5+1+1.5*\x,0.75);
}

\node[anchor=center] at (0.75,1){$=$};

\draw[pattern=north west lines, pattern color=red!70!blue!80, opacity=1, rounded corners = 2,thick](-0.625,1.5) rectangle ++(0.75,0.5);
\draw[pattern=crosshatch, pattern color=blue!80!green!80, opacity=1, rounded corners = 2,thick] (-1.125,0.75) rectangle ++(0.75,0.5);
\draw[fill=red!10!green!50!blue!20, opacity=1, rounded corners = 2,thick] (-0.625,0) rectangle ++(0.75,0.5);    
\foreach \x in {-1}
{
\draw[red!80!blue, line width=0.25mm, opacity=0.5] (1+1.5*\x,-0.25) -- (1+1.5*\x,0);
\draw[red!80!blue, line width=0.25mm, opacity=0.5] (1.5+1.5*\x,0.5) -- (1.5+1.5*\x,1.5);
\draw[red!80!blue, line width=0.25mm, opacity=0.5] (1+1.5*\x,2) -- (1+1.5*\x,2.25);
\draw[green!20!blue, line width=0.7mm, opacity=0.5] (1.5+1.5*\x,-0.25) -- (1.5+1.5*\x,0);
\draw[green!20!blue, line width=0.7mm, opacity=0.5] (1+1.5*\x,0.5) -- (1+1.5*\x,0.75);
\draw[green!20!blue, line width=0.7mm, opacity=0.5] (0.5+1.5*\x,1.25) -- (0.5+1.5*\x,2.25);
\draw[green!50!black, line width=1.15mm, opacity=0.75] (1.5+1.5*\x,2) -- (1.5+1.5*\x,2.25);
\draw[green!50!black, line width=1.15mm, opacity=0.75] (0.5+1.5*\x,-0.25) -- (0.5+1.5*\x,0.75);
\draw[green!50!black, line width=1.15mm, opacity=0.75] (1+1.5*\x,1.25) -- (1+1.5*\x,1.5);
}

\draw[black,thick] (5.325,1.375+0.0625) -- (5.325,2+0.0625);
\draw[black,thick] (5.675,1.375+0.0625) -- (5.675,2+0.0625);
\draw[fill=red!10!green!50!blue!20, opacity=1, rounded corners = 2,thick] (5.25,1.5+0.0625) rectangle ++(0.5,0.375);
\draw[black,thick] (5.325,0.625+0.0625) -- (5.325,1.25+0.0625);
\draw[black,thick] (5.675,0.625+0.0625) -- (5.675,1.25+0.0625);
\draw[fill=white, opacity=1, rounded corners = 2,thick] (5.25,0.75+0.0625) rectangle ++(0.5,0.375);
\draw[pattern=crosshatch, pattern color=blue!80!green!80, opacity=1, rounded corners = 2,thick] (5.25,0.75+0.0625) rectangle ++(0.5,0.375);
\draw[black,thick] (5.325,-0.125+0.0625) -- (5.325,0.5+0.0625);
\draw[black,thick] (5.675,-0.125+0.0625) -- (5.675,0.5+0.0625);
\draw[fill=white, opacity=1, rounded corners = 2,thick] (5.25,0+0.0625) rectangle ++(0.5,0.375);
\draw[pattern=north west lines, pattern color=red!70!blue!80, opacity=1, rounded corners = 2,thick] (5.25,0+0.0625) rectangle ++(0.5,0.375);
\node[anchor=west] at (6,1.75+0.0625){$U^{(\tau_{21})}$};
\node[anchor=west] at (6,1+0.0625){$U^{(\tau_{23})}$};
\node[anchor=west] at (6,0.25+0.0625){$U^{(\tau_{13})}$};
\node[anchor=west] at(5,2.375+0.0625){Gates:};
\node[anchor=west] at(5,-0.5+0.0625){$\tau_{ij}\equiv\tau_i-\tau_j$};

\end{tikzpicture}
\label{eq:braided-ybe}
\end{equation}
we can exchange the upper three layers of the circuit in Fig.~\hyperref[fig:three-circuits]{2(c)} with those of the circuit in Eq.~\eqref{eq:different-gate-circuit}, thus proving their equivalence.

The above identification of transfer matrices generating different-gate integrable circuits can be extended to larger numbers of inhomogeneities, as well as to circuits that are not space-periodic, i.e., circuits in which the system size $L$ is not a multiple of $q$. Having provided a proof-of-concept example, we instead proceed by highlighting how to quickly discern an integrable same-gate circuit from a given set of inhomogeneities.

\subsection{``Lattice-cut'' construction of periodic same-gate circuits} 
\label{sec:lattice-cut}

Given an arrangement of gates in a circuit, we have so far been able to determine the set of inhomogeneities $\bs{\nu}$ in the transfer matrix. A natural follow-up question is whether, for a given set of inhomogeneities, there exists an efficient method for determining the corresponding gate arrangement. Here, we will demonstrate a procedure which allows that for the same-gate circuits that are periodic in both space and time. Transfer matrices of such circuits have two different inhomogeneity parameters, i.e., $\nu_j\in\{\tau_1,\tau_2\}$ for $j=1,\ldots,L$, which form a periodic sequence $\bs{\nu}$.

\begin{figure}[ht!]
    \centering
    \hspace{1em}
    \includegraphics[width=0.9\textwidth]{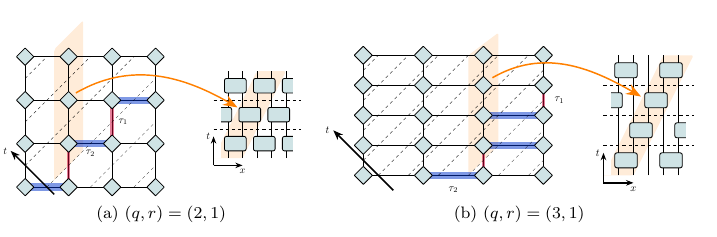}
    \vspace{-1em}
    \caption{Lattice cuts are the dashed lines which intersect vertical and horizontal lattice edges according to the ordering of $\tau_1$ and $\tau_2$ in the set of transfer-matrix inhomogeneities $\bs{\nu}$. We have highlighted $\tau_1$ (assigned to the vertical edges) with a thin red line, and $\tau_2$ (assigned to the horizontal edges) with a thick blue line. The lattice cuts correspond to time-slices in the quantum circuit---the direction of time is perpendicular to them. This scheme allows for a quick identification of the corresponding circuit.}
    \label{fig:circuit-q3r1}
\end{figure}

For clarity, we will consider two simple examples shown in Fig.~\ref{fig:circuit-q3r1}. In the first one the set of inhomogeneities $\bs{\nu}$ has a period $q=2$: we will assume it is a repetition of $(\tau_1,\tau_2)$. To deduce the corresponding circuit one imagines a rectangular lattice with gates on the vertices. The edges connecting the gates are assigned a parameter: $\tau_1$ for the vertical edges and $\tau_2$ for the horizontal ones. We now cut through the lattice with a line that intersects vertical and horizontal edges based on the order in which $\tau_{1}$ and $\tau_2$ appear in the sequence $\bs{\nu}$. The direction of time is perpendicular to the lattice cuts---each lattice cut represents a snapshot of the circuit at a particular discrete time. In this example we recognize a brickwork arrangement of gates---see Fig.~\hyperref[fig:circuit-q3r1]{3(a)}. In the second example we take a period $q=3$, the repeated ``unit cell'' in $\bs{\nu}$ being $(\tau_1,\tau_2,\tau_2)$. The lattice cut now periodically intersects a vertical edge, followed by two horizontal ones. Each lattice cut again corresponds to a particular discrete time, and the circuit obtained in this way now has $(q,r)=(3,1)$.

\section{Eigenstates and thermodynamic states}
\label{sec:eigenstates}

Our aim now shifts towards describing the eigenstates of the same-gate quantum unitary circuits that are periodic in space and time. As described in subsection~\ref{sec:lattice-cut}, such circuits can be identified via the ``lattice-cut'' construction, starting from a periodic set of transfer-matrix inhomogeneities $\nu_j\in\{\tau_1,\tau_2\}$,  where $\tau\equiv\tau_2-\tau_1$ is the gate parameter. 

Consider a circuit on $L\in q\mathbb{N}$ qubits, with space periodicity of $q$ sites and time periodicity of $q$ steps, for example, the circuits in Figs.~\hyperref[fig:circuits]{1(a)}--\hyperref[fig:circuits]{1(d)}. The set $\bs{\nu}$ of inhomogeneities in the transfer matrix~\eqref{eq:transfer-matrix} generating such a circuit has a period $q$: the inhomogeneities satisfy $\nu_{j+q}=\nu_j$, and we can identify a repeated subsequence $(\nu_1,\ldots,\nu_q)$ in $\bs{\nu}$, which we define as a \emph{unit cell}. We will assume that $\tau_{1,2}$ appears $n_{1,2}$-times in the unit cell, so that $n_1+n_2=q$. For example, the circuit in Fig.~\hyperref[fig:circuit-q3r1]{3(a)} has unit cell $(\tau_1,\tau_2)$ with $q=2$, $n_1=1$, and $n_2=1$, while the one in Fig.~\hyperref[fig:circuit-q3r1]{3(b)} has unit cell $(\tau_1,\tau_2,\tau_2)$ with $q=3$, $n_1=1$, and $n_2=2$. 

For the same-gate periodic circuits it generally holds that the $q$-step Floquet evolution operator $\mathds{U}$ is obtained from the transfer matrix~\eqref{eq:transfer-matrix} as
\begin{equation}
    \mathds{U}=\left[\mathds{T}(\tau_1)\right]^{-1}\mathds{T}(\tau_2),
    \label{eq:time-shift}
\end{equation}
while the $q$-site space shift towards the west (left) is
\begin{equation}
    \mathds{W}^q=\left[\mathds{T}(\tau_1)\right]^{n_1}\left[\mathds{T}(\tau_2)\right]^{n_2}.
    \label{eq:space-shift}
\end{equation}
Importantly, the way in which the time shift~\eqref{eq:time-shift} is expressed in terms of transfer matrices is the same for all same-gate circuits: it does not \emph{explicitly} depend on $n_1$ and $n_2$, as is the case for the space-shift operator~\eqref{eq:space-shift}.

\subsection{Eigenvalues of lattice-shift operators}

Due to Eqs.~\eqref{eq:time-shift} and~\eqref{eq:space-shift}, $\mathds{U}$ and $\mathds{W}^q$ are diagonal in the common eigenbasis of the commuting transfer matrices $\mathds{T}(\lambda)$. Its elements are parametrized by a set of $N$ rapidities $\bs{\lambda}=(\lambda_1,\ldots,\lambda_N)$, where $N$ is the number of quasiparticles---magnons. Provided that the rapidities satisfy Bethe equations, described in subsection~\ref{sec:bethe-eqs}, one has 
\begin{equation}
    \mathds{T}(\lambda)\ket{\bs{\lambda}}=\Lambda(\lambda;\bs{\lambda})\ket{\bs{\lambda}},
\end{equation}
where the eigenvalue
\begin{equation}
    \Lambda(\lambda;\bs{\lambda})=\prod_{j=1}^N \frac{\lambda\!-\!\lambda_j\!-\!\frac{\ii}{2}}{\lambda\!-\!\lambda_j\!+\!\frac{\ii}{2}}+\left[\!\left(\frac{\lambda\!-\!\tau_1}{\lambda\!-\!\tau_1\!+\!\ii}\right)^{n_1} \left(\frac{\lambda\!-\!\tau_2}{\lambda\!-\!\tau_2\!+\!\ii}\right)^{n_2}\right]^{\frac{L}{q}}\prod_{j=1}^N \frac{\lambda\!-\!\lambda_j\!+\!\frac{3\ii}{2}}{\lambda\!-\!\lambda_j\!+\!\frac{\ii}{2}}
    \label{eq:T-eigenvalue}
\end{equation}
is obtained following the procedure known as algebraic Bethe ansatz---see, e.g., Ref.~\cite{faddeev1996}.\footnote{Setting $\tau_1$ and $\tau_2$ to zero, Eq.~\eqref{eq:T-eigenvalue} correctly reproduces the transfer-matrix eigenvalue in the homogeneous isotropic spin-$\frac{1}{2}$ Heisenberg model. Compared to Ref.~\cite{faddeev1996}, our spectral parameter $\lambda$ is shifted for $\frac{\ii}{2}$, and our eigenvalue is normalized such that there is no prefactor in the first term (normalization ensures unitarity of the gate $U$).} Due to the normalization of the $R$ matrix, only the prefactor in the second term contains $\tau_1$ and $\tau_2$. 

The eigenvalues of the lattice-shift operators~\eqref{eq:time-shift} and~\eqref{eq:space-shift} are expressed in terms of $\Lambda(\lambda;\bs{\lambda})$ evaluated at $\lambda=\tau_1,\tau_2$. In these two points, the second term in Eq.~\eqref{eq:T-eigenvalue} disappears, such that the eigenvalue $\Lambda(\tau_{1,2};\bs{\lambda})$ no longer depends on $n_1$ and $n_2$. We are then left with
\begin{equation}
    \mathds{U}\ket{\bs{\lambda}}=\prod_{j=1}^N\left(\frac{\lambda_j\!-\!\tau_1\!-\!\frac{\ii}{2}}{\lambda_j\!-\!\tau_1\!+\!\frac{\ii}{2}}\right)\left(\frac{\lambda_j\!-\!\tau_2\!+\!\frac{\ii}{2}}{\lambda_j\!-\!\tau_2\!-\!\frac{\ii}{2}}\right)\ket{\bs{\lambda}}
    \label{eq:U-eigenvalues}
\end{equation}
and
\begin{equation}
   \mathds{W}^q\ket{\bs{\lambda}}=\prod_{j=1}^N \left(\frac{\lambda_j\!-\!\tau_1\!+\!\frac{\ii}{2}}{\lambda_j\!-\!\tau_1\!-\!\frac{\ii}{2}}\right)^{n_1} \left(\frac{\lambda_j\!-\!\tau_2\!+\!\frac{\ii}{2}}{\lambda_j\!-\!\tau_2\!-\!\frac{\ii}{2}}\right)^{n_2}\ket{\bs{\lambda}}.
    \label{eq:S-eigenvalues}
\end{equation}
The dependence on $n_1$ and $n_2$ in Eq.~\eqref{eq:S-eigenvalues} is therefore due to the explicit dependence of $\mathds{W}^q$ on these two parameters---see Eq.~\eqref{eq:space-shift}.

Since $\mathds{U}$ and $\mathds{W}^q$ are lattice shifts for $q$ sites in time and in space, respectively, we can express their eigenvalues in terms of magnon \emph{quasienergies} and \emph{quasimomenta}. In particular, Eqs.~\eqref{eq:U-eigenvalues} and~\eqref{eq:S-eigenvalues} suggest that we can write
\begin{equation}
    \mathds{U}\ket{\bs{\lambda}}=\e^{\ii q\sum_{j=1}^N\varepsilon_1(\lambda_j)}\ket{\bs{\lambda}}
\end{equation}    
and    
\begin{equation}
    \mathds{W}^q\ket{\bs{\lambda}}=\e^{-\ii q \sum_{j=1}^N p_1(\lambda_j)}\ket{\bs{\lambda}},
\end{equation}
where the quasienergy reads
\begin{equation}
    \varepsilon_1(\lambda)=\frac{1}{q}\,k_1(\lambda\!-\!\tau_1)-\frac{1}{q}\,k_1(\lambda\!-\!\tau_2),
    \label{eq:quasienergy}
\end{equation}
while
\begin{equation}
    p_1(\lambda)=\frac{n_1 }{q}\,k_1(\lambda\!-\!\tau_1)+\frac{n_2}{q}\, k_1(\lambda\!-\!\tau_2)
    \label{eq:quasimomentum}
\end{equation}
is the quasimomentum. Here, $k_1(\lambda)\equiv\ii \log[(\lambda+\frac{\ii}{2})/(\lambda-\frac{\ii}{2})]$ denotes the quasimomentum carried by a single magnon in the isotropic spin-$\frac{1}{2}$ Heisenberg model. The number of terms containing $\tau_1$ and $\tau_2$ in the expression for the quasimomentum $p_1(\lambda)$, respectively, $n_1$ and $n_2$, reflects the pattern of inhomogeneities in the transfer matrix generating the circuit. On the other hand, the expression for the quasienergy is independent of $n_1$ and $n_2$, consistently with our previous observation that $\mathds{U}$ does not explicitly depend on these two parameters.

\subsection{Bethe equations}
\label{sec:bethe-eqs}

Rapidities that specify the eigenstates of the transfer matrix~\eqref{eq:transfer-matrix} are obtained from the quantization condition (Bethe equations)
\begin{equation}
    \e^{\ii L p_1(\lambda_j)}\prod_{\ell=1 \atop \ell\ne j}^N \mathcal{S}(\lambda_j-\lambda_\ell)=1,\qquad j=1,\ldots,N,
    \label{eq:quantization-cond}
\end{equation}
where the quasimomentum $p_1(\lambda)$ is defined in Eq.~\eqref{eq:quasimomentum}, and 
\begin{equation}
    \mathcal{S}(\Delta\lambda)=\frac{\Delta\lambda+\ii}{\Delta\lambda-\ii}    
\end{equation}
is the scattering phase between two quasiparticles whose rapidities differ for $\Delta\lambda$. Explicitly, the Bethe equations~\eqref{eq:quantization-cond} read
\begin{equation}
    \left(\frac{\lambda_j-\tau_1+\frac{\ii}{2}}{\lambda_j-\tau_1-\frac{\ii}{2}}\right)^{L_1}\left(\frac{\lambda_j-\tau_2+\frac{\ii}{2}}{\lambda_j-\tau_2-\frac{\ii}{2}}\right)^{L-L_1}=\prod_{\ell=1\atop \ell\ne j}^N\left(\frac{\lambda_j-\lambda_\ell+\ii}{\lambda_j-\lambda_\ell-\ii}\right),
    \label{eq:bethe-eqs}
\end{equation}
where $L_1\equiv Ln_1/q$ is the total number of inhomogeneities $\tau_1$ in the entire transfer matrix. The latter number is the SEC (spectral equivalence class) classification number introduced in subsection~\ref{sec:integrable-geometries}, i.e.,  $L_1\equiv p$.\footnote{In order not to confuse the SEC number $p$ with the quasimomentum, we will denote it by $L_1$ from now on.} 

Bethe equations~\eqref{eq:bethe-eqs} imply that, in a system of size $L$, with fixed $\tau_1$ and $\tau_2$, the rapidities that characterize the spectrum of the circuit with a period $q$ depend only on $L_1$. Here, the period can even  be $q=L$ (i.e., no nontrivial translational invariance), in which case $L_1=n_1$. Indeed, Bethe equations remain unaffected by an arbitrary redistribution of the inhomogeneities among the $R$ matrices that form the transfer matrix~\eqref{eq:transfer-matrix}. In particular, all integrable circuits originating in a transfer matrix that contains $L_1$ inhomogeneity parameters $\tau_1$ (the rest being $\tau_2$), irrespective of whether the circuit is periodic or not, will have the same spectrum of quasienergies, and will thus belong to the same SEC.

\subsection{Description of a thermodynamic state}
\label{sec:thermodynamic-content}

We assume that the solution $\bs{\lambda}$ of Bethe equations organizes into {\em strings}---subsets of rapidities that extend in the imaginary direction of a complex plane. The same assumption is usually made when solving the Bethe equations of continuous-time Heisenberg model, in which case $\tau_{1,2}=0$. Neglecting the $O(\e^{-L})$ corrections, the rapidities forming an $m$-string ($m\in\mathbb{N}$) read
\begin{equation}
    \lambda^{m,j}_\alpha = \lambda^m_\alpha+\frac{\ii}{2}(m+1-2j),\qquad j=1,\ldots,m.
\end{equation}
Here, $\alpha$ labels the string and $\lambda_\alpha^m\in\mathbb{R}$ is its real {\em center}. Such an $m$-string corresponds to $m$ magnons bound by the attractive interaction. Bound together, these magnons carry quasienergy $\varepsilon_m(\lambda_{\alpha}^{m})=\sum_{j=1}^m \varepsilon_1(\lambda_\alpha^{m,j})$ and quasimomentum $p_m(\lambda_\alpha^{m})=\sum_{j=1}^m p_1(\lambda_\alpha^{m,j})$ which generalize those of a single quasiparticle, given in Eqs.~\eqref{eq:quasienergy} and~\eqref{eq:quasimomentum}, respectively. In particular, we have
\begin{equation}
    \varepsilon_m(\lambda)=\frac{1}{q}\,k_m(\lambda-\tau_1)-\frac{1}{q}\,k_m(\lambda-\tau_2),
    \label{eq:string-quasienergy}
\end{equation}
and
\begin{equation}
    p_m(\lambda)=\frac{n_1}{q} \, k_m(\lambda\!-\!\tau_1)\!+\!\frac{n_2}{q} \, k_m(\lambda\!-\!\tau_2),
    \label{eq:string-quasimomentum}
\end{equation}
where
\begin{equation}
    k_m(\lambda)\equiv \ii \log\left(\frac{\lambda+m \frac{\ii}{2}}{\lambda-m \frac{\ii}{2}}\right)
    \label{eq:heisenberg-momentum}
\end{equation}
denotes the $m$-string quasimomentum in the isotropic homogeneous spin-$\frac{1}{2}$ Heisenberg model, in which $\tau_{1,2}=0$. A single quasiparticle can be thought of as a $1$-string [cf. Eqs.~\eqref{eq:quasienergy},~\eqref{eq:quasimomentum}].

In the thermodynamic limit, string centers densely populate the real axis in a complex plane. A thermodynamic state of the system is then specified by the distributions of the occupied string centers 
\begin{equation}
    \rho^{}_m(\lambda)\equiv n^{}_m(\lambda)\rho^{\rm (tot)}_m(\lambda),\quad m\in\mathbb{N} \qquad \Longrightarrow\qquad \bs{\rho}=\mathbf{n}\bs{\rho}^{\rm(tot)},
    \label{eq:root-density}
\end{equation}
where the following objects have appeared: 
\begin{itemize}
    \item[--] the total state density $\rho_m^{\rm (tot)}(\lambda)$, defined such that $L\rho_m^{\rm (tot)}(\lambda){\rm d}\lambda$ is the total number of available string centers in the rapidity interval $[\lambda,\lambda+{\rm d}\lambda)$;
    \item[--] the occupancy function $n_m(\lambda)$, determining the ratio of the occupied string centers in $[\lambda,\lambda+{\rm d}\lambda)$;
    \item[--] vectors $\bs{\rho}$ and $\bs{\rho}^{\rm (tot)}$, whose components, $\rho_m(\lambda)$ and $\rho_m^{\rm (tot)}(\lambda)$, respectively, have a discrete index $m$ and a continuous-one $\lambda$;
    \item[--] a diagonal matrix $\mathbf{n}$ with elements $\delta_{\ell,m}\delta(\lambda-\eta)n_m(\lambda)$.
\end{itemize}
In the compact vector notation on the right-hand side in Eq.~\eqref{eq:root-density}, matrix product should be interpreted as a summation over the discrete index and a convolution over the continuous-one.

Bethe equations can be reduced to equations involving the real string centers only~\cite{takahashi1999}. In particular, in the thermodynamic limit, they become the Bethe-Yang equations, compactly written as
\begin{equation}
    (\mathbf{1}+\mathbf{Kn})\bs{\rho}^{\rm (tot)}=\frac{1}{2\pi}\bs{p}',
    \label{eq:Bethe-Yang-eq}
\end{equation}
in which $\bs{p}'$ denotes a vector of momentum derivatives $p_m'(\lambda)\equiv{\rm d}p_m/{\rm d}\lambda$, $\mathbf{1}$ is a diagonal matrix with elements $\delta_{\ell,m}\delta(\lambda-\eta)$, and the entries of matrix $\mathbf{K}$ read
\begin{equation}
    K_{\ell,m}(\lambda-\eta)=\sum_{j=|\ell-m|/2}^{(\ell+m)/2-1}\frac{1}{2\pi}\left[k'_{2j}(
    \lambda-\eta)+k'_{2j+2}(\lambda-\eta)\right],
\end{equation}
with $k_m(\lambda)$ given in Eq.~\eqref{eq:heisenberg-momentum}.

Since $\mathbf{n}$ and $\boldsymbol{\rho}^{\rm (tot)}$ are both unknowns, Eq.~\eqref{eq:Bethe-Yang-eq} has to be complemented by another equation, obtained using the maximum entropy principle. Specifically, one has to maximize the thermodynamic entropy, which counts the microstates described by the same pair $(\mathbf{n},\boldsymbol{\rho}^{\rm (tot)})$, constrained by the conserved charges~\cite{yang1969}. In the remainder of this paper, we will be interested in the simplest nontrivial thermodynamic state of a U(1)-invariant quantum unitary circuit---the infinite-temperature grand-canonical Gibbs ensemble at a chemical potential $\mu$, whose density matrix is
\begin{equation}
    \varrho_\mu=\frac{\e^{-\mu {\rm S}^\zz}}{{\rm Tr}(\e^{-\mu {\rm S}^\zz})}.
    \label{eq:density-matrix}
\end{equation} 
Combined with the Bethe-Yang equation~\eqref{eq:Bethe-Yang-eq}, the maximum entropy principle yields
\begin{equation}
    \log\left(n_m(\lambda)^{-1}-1\right)=\mu m-\sum_\ell \int{\rm d}\eta\, K_{m,\ell}(\lambda-\eta)\log\left[1-n_\ell(\eta)\right].
    \label{eq:gc-occupancy-eq}
\end{equation}
This equation in turn admits the following solution for the occupancy ratio~\cite{takahashi1999}:
\begin{equation}
    n_m=\kar_m^{-2},\qquad \kar_m\equiv\frac{\sinh\left([m+1]\frac{\mu}{2}\right)}{\sinh\left(\frac{\mu}{2}\right)},\qquad (m\in\mathbb{N}).
\end{equation}

Crucially, the occupancy ratio $n_m$ is independent of $\lambda$, and thus of $\tau_{1,2}$. Bethe-Yang equation~\eqref{eq:Bethe-Yang-eq} is therefore a linear equation for $\bs{\rho}^{\rm (tot)}$. Since, according to Eq.~\eqref{eq:string-quasimomentum}, $\bs{p}'$ is a sum of contributions from two homogeneous Heisenberg models with $\tau_{1,2}$, respectively, $\bs{\rho}^{\rm (tot)}$ will also be a sum of two total state densities~\cite{zadnikQuantumManybodySpin2024a}. In particular,
\begin{equation}
    \rho_m^{\rm (tot)}(\lambda)=\frac{n_1}{q}\rho_m^{\rm Heis}(\lambda-\tau_1)+\frac{n_2}{q}\rho_{m}^{\rm Heis}(\lambda-\tau_2),
    \label{eq:rho-solution}
\end{equation}
where
\begin{equation}
    \rho_m^{\rm Heis}(\lambda)\equiv\frac{\kar_m}{2\pi\kar_1}\left(\frac{k'_m(\lambda)}{\kar_{m-1}}-\frac{k'_{m+2}(\lambda)}{\kar_{m+1}}\right)
\end{equation}
solves the Bethe-Yang equation~\eqref{eq:Bethe-Yang-eq} specialized to a (continuous-time) spin-$\frac{1}{2}$ Heisenberg model with $\tau_{1,2}=0$, i.e., with a homogeneous transfer matrix, and with $p_m(\lambda)=k_m(\lambda)$.\footnote{Specifically, the Bethe-Yang equation specialized to the (continuous-time) Heisenberg model with $\tau_{1,2}=0$ reads $(\mathbf{1}+\mathbf{Kn})\bs{\rho}^{\rm Heis}=\frac{1}{2\pi}\bs{k}'$.} This solution has been computed in Ref.~\cite{ilievski2018} and later generalized to arbitrary integer or half-integer spin in Ref.~\cite{zadnikQuantumManybodySpin2024a}. 

It is worthwile noting that, by Eq.~\eqref{eq:rho-solution},  the number of vacancies $L\rho_m^{\rm(tot)}(\lambda){\rm d}\lambda$ in the spectral parameter interval $[\lambda,\lambda+{\rm d}\lambda)$ involves contributions from two sublattices of sizes $L_1=Ln_1/q$ and $L_2=Ln_2/q$, respectively. Indeed, $L_i\rho_m^{\rm Heis}(\lambda-\tau_i){\rm d}\lambda$, where $i\in\{1,2\}$, are the numbers of vacancies in the shifted intervals $[\lambda-\tau_i,\lambda-\tau_i+{\rm d}\lambda)$: they count the vacancies in the grand-canonical state of a Heisenberg model on the respective sublattice. With this remark, and having obtained $n_m$ and $\rho_m^{\rm (tot)}(\lambda)$, we conclude the description of the thermodynamic state. 

\section{Charge transport in asymmetric circuit geometries}
\label{sec:hydro}

We will now study the transport properties of our integrable quantum unitary circuits. Of particular interest is the transport of their U(1) charge (magnetization)
\begin{equation}
    {\rm S}^\zz=\sum_{j=1}^L {\rm s}^\zz_j,\qquad   \left({\rm s}^\zz_j=\frac{1}{2}\sigma_j^{\zz}\right),
\end{equation}
characterized by the asymptotic behaviour of the dynamical spin susceptibility~\cite{doyon2017,denardis2019,denardis2022}
\begin{equation}
    S(x,t)=\langle {\rm s}^\zz(x,t){\rm s}^\zz(0,0)\rangle^c_{\mu}.
    \label{eq:spin-susceptibility}
\end{equation}
Here, $\langle AB\rangle^c_{\mu}\equiv \langle AB\rangle_\mu-\langle A\rangle_\mu\langle B\rangle_\mu$ denotes the connected correlation function in the grand-canonical ensemble~\eqref{eq:density-matrix} with chemical potential $\mu$, and coordinates $x,t$ are coarse-grained over a period $q$ in space and in time: 
\begin{equation}
    {\rm s}^\zz(x,t)\equiv\mathds{U}^{-t}\left(\frac{1}{q}\sum_{\ell=1}^q{\rm s}^\zz_{q(x-1)+\ell}\right)\mathds{U}^{t}.
    \label{eq:coarse-grained-spin}
\end{equation}
The coordinate $x$ can moreover be shifted so that $x=0$ corresponds to the center of the system---we will implicitly assume this convention from now on.

\begin{figure}[ht!]
\centering
\begin{tikzpicture}
    \node[anchor=south west,inner sep=0] (image) at (0,0) {\includegraphics[width=0.8\textwidth]{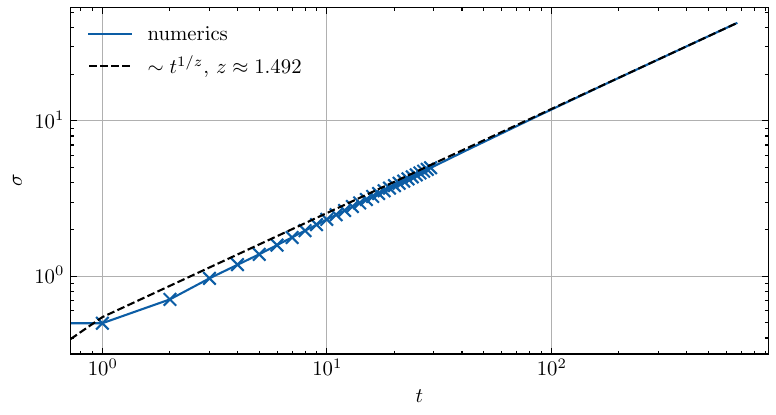}};
    \begin{scope}[x={(image.south east)},y={(image.north west)}]
        \node[anchor=south west,inner sep=0] (image) at (0.45,0.15) {\includegraphics[width=0.42\textwidth]{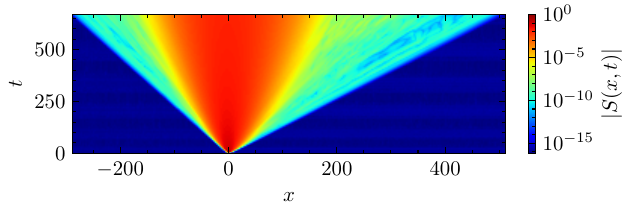}};
    \end{scope}
\end{tikzpicture}
\caption{Numerically obtained dynamical exponent $z$, Eq.~\eqref{eq:dynamical-exponent}, at half-filling ($\mu = 0$), for the isotropic Heisenberg gate with $\tau=1$, arranged in the asymmetric $(q, r) = (3, 1)$ geometry depicted in Fig.~\hyperref[fig:circuit-q3r1]{3(b)}. The dynamical exponent is close to the expected one, $z = 3/2$. The main diagram shows the linear fit to the time-dependent variance $\sigma$ for large times (the fitted curve overlaps with the numerical data at later times). The inset depicts the dynamical spin susceptibility. The circuit consists of $L=4008$ qubits. Calculations were performed using a TEBD algorithm with maximal bond dimension $d_{\rm max}=256$. We have used the linear-response approximation, Eq.~\ref{eq:lin_resp}, with $\Delta \mu = 10^{-4}$.}
\label{fig:transport_numerics1}
\end{figure}

The large-scale behavior of the dynamical spin susceptibility $S(x,t)$ is captured by its hydrodynamic mode decomposition~\cite{doyon2017}, which will be considered in subsection~\ref{sec:susceptibility-hydro-decomposition}, and whose essential ingredient is the quasiparticle content of the thermodynamic state (see subsection~\ref{sec:thermodynamic-content}). Here, we will use it to determine how the circuit geometry affects the U(1)-charge transport. We are motivated by the fact that a broken space-reflection symmetry $\mathcal{P}$ may lead to a drift in $S(x,t)$, signaled by a nonzero {\em drift velocity}
\begin{equation}
    v_{\rm d}=\frac{1}{\chi}\lim_{t\to\infty}\int{\rm d}x \,\frac{x}{t}\, S(x,t),
    \label{eq:drift-velocity}
\end{equation}
where $\chi=\int{\rm d}x \, S(x,t)$ is the {\em static spin susceptibility}. In particular, nonzero drift velocity has recently been reported in circuits, in which each unitary gate breaks the space-reflection symmetry by acting on two different degrees of freedom~\cite{gong2022,zadnikQuantumManybodySpin2024a,richelli2024}. In quantum circuits considered herein, all degrees of freedom are the same, and the unitary gates are thus symmetric. However, the space-reflection symmetry can nevertheless be broken by the gates' arrangement, an example being shown in Fig.~\hyperref[fig:circuit-q3r1]{3(b)}. In order to characterize the effects of the left-right asymmetry (i.e., of the broken $\mathcal{P}$), we will consider not only the drift velocity of the dynamical spin susceptibility, but also its higher time-scaled centered moments of odd order, i.e.,
\begin{equation}
    s^{(n)}\equiv \lim_{t\to\infty}\int {\rm d}x \left(\frac{x}{t}-v_{\rm d}\right)^n S(x,t),
    \label{eq:scaled-moments}
\end{equation}
for odd $n\in\mathbb{N}$, $n>1$. Before proceeding, we remark that, for general $n$ (odd or even), the scaled centered moments $s^{(n)}$ describe the spreading of charge correlations in the moving frame.\footnote{Note that, due to Eq.~\eqref{eq:drift-velocity}, the first scaled centered moment vanishes trivially, $s^{(1)}=0$.} In particular, the second centered scaled moment is referred to as the Drude weight, $D\equiv s^{(2)}$. If nonzero, it implies ballistic spin transport. More generally, the growth of the susceptibility's variance at large times defines the dynamical exponent $z$ through
\begin{equation}
    \sigma^2(t)\equiv\int{\rm d}x (x-v_{\rm d}t)^2 S(x,t)\simeq t^{2/z}.
    \label{eq:dynamical-exponent}
\end{equation}
For example, $z=1$ is typical for ballistic transport with a finite Drude weight $D$, while diffusive scaling is signaled by $z=2$. At half-filling ($\mu=0$) the SU(2) symmetry of the unitary gate and the state kicks in. It has been shown that the dynamical exponent is then fractional, specifically, $z=\frac{3}{2}$~\cite{ilievski2021}. We confirm this numerically also for circuits that are not of the brickwork type---see Fig.~\ref{fig:transport_numerics1}.

\subsection{Dynamical spin susceptibility on ballistic scale}
\label{sec:susceptibility-hydro-decomposition}

The Euler-scale behaviour of the dynamical spin susceptibility~\eqref{eq:spin-susceptibility} is captured by its decomposition in terms of ballistically propagating modes~\cite{doyon2017},
\begin{equation}
    S(x,t)\simeq \sum_m\int{\rm d}\lambda \, \delta\!\left(x-v^{\rm eff}_m(\lambda)t\right) \rho_m^{\rm (tot)}(\lambda)n_m(1-n_m) (q^{\rm dr}_m)^2.
    \label{eq:hydro-susceptibility}
\end{equation}
Here, $q_m^{\rm dr}$ is the \emph{dressed magnetization} carried by a quasiparticle of $m$-th species, propagating with an effective mode velocity
\begin{equation}
    v^{\rm eff}_m(\lambda)=\frac{(\varepsilon_m')^{\rm dr}(\lambda)}{(p_m')^{\rm dr}(\lambda)}.
    \label{eq:effective-velocity}
\end{equation}
Dressed quantities $\boldsymbol{q}^{\rm dr}$ are defined as solutions of the equation
\begin{equation}
    (\mathbf{1}+\mathbf{Kn})\boldsymbol{q}^{\rm dr}=\boldsymbol{q}.
    \label{eq:dressing}
\end{equation}
To obtain the explicit expressions for $v^{\rm eff}_m$ and $q^{\rm dr}_m$, respectively, we note the following:
\begin{enumerate}
    \item[(1)] Through Bethe-Yang equation~\eqref{eq:Bethe-Yang-eq}, the total state density corresponds to a dressed momentum derivative, $2\pi\boldsymbol{\rho}^{\rm (tot)}=(\boldsymbol{p}')^{\rm dr}$. Specializing to $\tau_{1,2}=0$, this becomes $2\pi \bs{\rho}^{\rm Heis}= (\bs{k}')^{\rm dr}$, and using it in the quasienergy~\eqref{eq:string-quasienergy} and quasimomentum~\eqref{eq:string-quasimomentum}, we find 
    \begin{equation}
        v^{\rm eff}_m(\lambda)=\frac{\rho_m^{\rm Heis}(\lambda-\tau_1)-\rho_m^{\rm Heis}(\lambda-\tau_2)}{n_1\,\rho^{\rm Heis}_m(\lambda-\tau_1)+n_2\,\rho^{\rm Heis}_m(\lambda-\tau_2)}.
    \end{equation}
    \item[(2)] Differentiating Eq.~\eqref{eq:gc-occupancy-eq} on the chemical potential $\mu$ yields an equation of the form~\eqref{eq:dressing} in which we recognize the dressing of the U(1) charge:
    \begin{equation}
        q_m=m \quad \longrightarrow \quad q^{\rm dr}_m=\partial_\mu\log(n_m^{-1}-1).
    \end{equation}
\end{enumerate}
Along with the results of subsection~\ref{sec:thermodynamic-content}, we now have all of the ingredients to compute the time-scaled centered moments~\eqref{eq:scaled-moments} of the dynamical spin susceptibility.

\subsubsection{Zero drift velocity.} We first use the hydrodynamic decomposition~\eqref{eq:hydro-susceptibility} in the expression for the drift velocity~\eqref{eq:drift-velocity}, to obtain
\begin{equation}
    v_{\rm d}=\frac{\sum_m n_m(1\!-\!n_m)(q^{\rm dr}_m)^2\left[\int{\rm d}\lambda\,\rho_m^{\rm Heis}(\lambda\!-\!\tau_1)-\int{\rm d}\lambda\,\rho_m^{\rm Heis}(\lambda\!-\!\tau_2)\right]}{\sum_m n_m(1\!-\!n_m) (q^{\rm dr}_m)^2\left[n_1 \int{\rm d}\lambda\,\rho^{\rm Heis}_m(\lambda\!-\!\tau_1)+n_2 \int{\rm d}\lambda\,\rho^{\rm Heis}_m(\lambda\!-\!\tau_2)\right]}=0.
    \label{eq:zero-drift}
\end{equation}
The contributions from two sublattices (one with $\tau_1$, the other with $\tau_2$) cancel out in the numerator, resulting in zero drift. Crucially, this holds irrespective of how we choose the numbers $n_{1,2}$ of the inhomogeneities $\tau_{1,2}$ in the spatial period of the transfer matrix~\eqref{eq:transfer-matrix}. In particular, setting $n_1 \ne n_2$ we obtain a quantum circuit with a broken space-reflection symmetry: since $n_1+n_2=q$ is the period of the circuit, differing $n_1$ and $n_2$ necessarily lead to an asymmetric arrangement of quantum gates. The result~\eqref{eq:zero-drift} therefore implies that a left-right asymmetry due to the geometry of the circuit is not sufficient for the drift to appear in the dynamical spin susceptibility. 

The reason for this lies in the propagator, which always expresses as $\mathds{U}=[\mathds{T}(\tau_1)]^{-1}\mathds{T}(\tau_2)$, and therefore does not explicitly depend on $n_1$ and $n_2$. Since all spins in the lattice are the same, $s=\frac{1}{2}$, the quasienergy obtained from the eigenvalue of $\mathds{U}$ is always a difference of contributions from the same type of quasiparticles---see Eq.~\eqref{eq:string-quasienergy}. In order for the numerator in Eq.~\eqref{eq:zero-drift} to be nonzero, one should therefore require different types of quasiparticle contributions from the two sublattices of respective sizes $L_1=L n_1/q$ and $L_2=L n_2/q$. This can be achieved by changing the degrees of freedom on one of the two sublattices, which indeed leads to a drift in the dynamics~\cite{gong2022,zadnikQuantumManybodySpin2024a,richelli2024}.

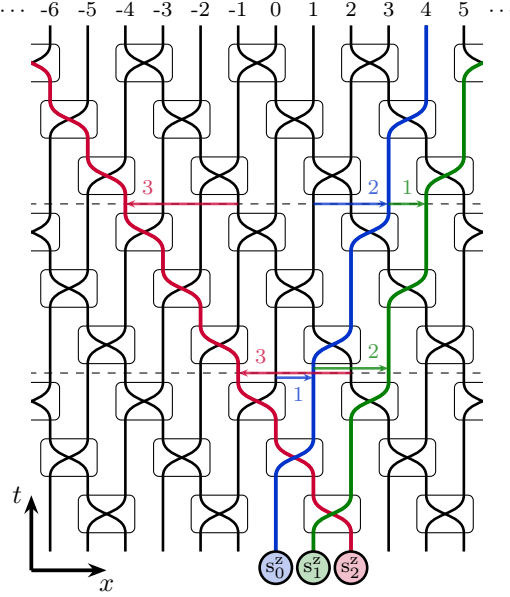
\begin{figure}[ht!]
\centering
\begin{tikzpicture}[baseline=(current  bounding  box.center),scale=1]
\foreach \x in {0,0.5,...,5.5}{
    \draw[line width=0.35mm] (0.5+\x, -6.25) -- (0.5+\x, 0.75);
}

\foreach \t in {0, 1, 2}{
    \foreach \x in {0,1.5,...,6}{
        \foreach \l in {0, 1, 2} {
            \draw[fill=white, rounded corners=2] (-0.125 + \x + 0.5*\l,-0.75*\l - 2.25*\t) rectangle ++(0.75,0.5);
            \draw[line width=0.35mm] (\x + 0.5*\l, -0.75*\l - 2.25*\t) to[out=90, in=-90] (\x+0.5 +0.5*\l, 0.5 - 0.75*\l - 2.25*\t);
            \draw[line width=0.35mm] (\x + 0.5 + 0.5*\l, -0.75*\l - 2.25*\t) to[out=90, in=-90] (\x +0.5*\l, 0.5 - 0.75*\l - 2.25*\t);
        }
    }
    \ifnum \t < 2
        \draw[dashed] (0, -1.625 - 2.25*\t) -- (6.5, -1.625 - 2.25*\t);
    \fi
}

\foreach \x in {0, 1, ..., 8} {
    \draw[red!80!blue, line width=0.5mm] (4.5 - 0.5*\x, -6.25 + 0.75*\x) -- (4.5 - 0.5*\x, -6 + 0.75*\x);
    \draw[red!80!blue, line width=0.5mm] (4.5 -0.5*\x, -6 + 0.75*\x) to[out=90, in=-90] (4 - 0.5*\x, -5.5 + 0.75*\x);
}
\foreach \x in {0, 1, ..., 4} {
    \ifnum \x > 0
        \draw[green!50!black, line width=0.5mm] (4 + 0.5*\x, -7 + 1.5*\x) -- (4 + 0.5*\x, -6 + 1.5*\x);
    \else
        \draw[green!50!black, line width=0.5mm] (4, -6.25) -- (4, -6);
    \fi
    \draw[green!50!black, line width=0.5mm] (4 + 0.5*\x, -6 + 1.5*\x) to[out=90, in=-90] (4.5 + 0.5*\x, -5.5 + 1.5*\x);
}
\foreach \x in {0, 1, ..., 4} {
    \draw[green!20!blue, line width=0.5mm] (3.5 + 0.5*\x, -6.25 + 1.5*\x) -- (3.5 + 0.5*\x, -5.25 + 1.5*\x);
    \ifnum \x < 4
        \draw[green!20!blue, line width=0.5mm] (3.5 + 0.5*\x, -5.25 + 1.5*\x) to[out=90, in=-90] (4 + 0.5*\x, -4.75 + 1.5*\x);
    \fi
}

\fill[white] (-1.25,-1.75 - 5) rectangle ++(1.5,7.5);
\fill[white] (6.25,-1.75 - 5) rectangle ++(1.5,7.5);

\fill[green!20!blue, opacity=0.25] (3.5, -6.465) circle (0.215);
\draw[line width=0.35mm] (3.5, -6.465) circle (0.215) node {\footnotesize{${\rm s}^\zz_0$}};
\fill[green!50!black, opacity=0.25] (4, -6.465) circle (0.215);
\draw[line width=0.35mm] (4, -6.465) circle (0.215) node {\footnotesize{${\rm s}^\zz_1$}};
\fill[red!80!blue, opacity=0.25] (4.5, -6.465) circle (0.215);
\draw[line width=0.35mm] (4.5, -6.465) circle (0.215) node {\footnotesize{${\rm s}^\zz_2$}};

\draw[-{Stealth[scale=0.5]}, red!80!blue, line width=0.35mm, opacity=0.75] (4.5, -3.875) -- (3, -3.875) node [above, pos=0.8] {\footnotesize{$3$}};
\draw[-{Stealth[scale=0.5]}, green!20!blue, line width=0.35mm, opacity=0.75] (3.5, -3.9375) -- (4, -3.9375) node [below, pos=0.6] {\footnotesize{$1$}};
\draw[-{Stealth[scale=0.5]}, green!50!black, line width=0.35mm, opacity=0.75] (4, -3.8125) -- (5, -3.8125) node [above, pos=0.8] {\footnotesize{$2$}};

\draw[-{Stealth[scale=0.5]}, red!80!blue, line width=0.35mm, opacity=0.75] (3, -1.625) -- (1.5, -1.625) node [above, pos=0.8] {\footnotesize{$3$}};
\draw[-{Stealth[scale=0.5]}, green!20!blue, line width=0.35mm, opacity=0.75] (4, -1.625) -- (5, -1.625) node [above, pos=0.8] {\footnotesize{$2$}};
\draw[-{Stealth[scale=0.5]}, green!50!black, line width=0.35mm, opacity=0.75] (5, -1.625) -- (5.5, -1.625) node [above, pos=0.5] {\footnotesize{$1$}};

\draw[-{Stealth[scale=0.75]}, black, line width=0.5mm] (0+0.25,-6.5) -- (0+0.25,-5.5);
\draw[-{Stealth[scale=0.75]}, black, line width=0.5mm] (0+0.25,-6.5) -- (1+0.25,-6.5);
\node[anchor=east] at (0+0.25,-5.5){$t$};
\node[anchor=north] at (1+0.25,-6.5){$x$};
\foreach \x in {-6,...,5}
{
\node[anchor=south] at (0.5*\x + 3.5,0.75){\footnotesize{\x}};
}
\node[anchor=south] at (0,0.75){\footnotesize{$\cdots$}};
\node[anchor=south] at (6.5,0.75){\footnotesize{$\cdots$}};
\end{tikzpicture}
\caption{Sketch of ${\rm
s}^\zz(0,t)=\mathds{U}^{-t}\left(\frac{1}{q}\sum_{\ell=1}^q{\rm
s}^\zz_{\ell}\right)\mathds{U}^{t}$ in the $(q, r) = (3, 1)$ geometry with permuting gates $U = P$. The coordinate $x$ is shifted such that $x=0$ is in the center of the system. The crossing lines denote the permutations (in the Heisenberg picture), and
the colours denote the propagation of one-site operators ${\rm s}^\zz$. Since the observables are evolved in the Heisenberg picture, the
propagation is tilted in the direction opposite to the one shown in
Fig.~\ref{fig:circuit-q3r1}.}
\label{fig:perm}
\end{figure}

Zero drift can be more intuitively understood in the limit $\tau \to
\infty$, in  which the unitary gate becomes a permutation, $U = P$. Since $P^{}_{j, j+ 1} {\rm s}^\zz_{j} P^{}_{j, j+1} = {\rm s}^\zz_{j + 1}$,
the coarse-grained spin operators ${\rm s}^\zz(x, t)$ propagate on the edges of the light-cone of the circuit.\footnote{This is also precisely where one expects nonzero correlations between one-site local observables in dual-unitary circuits~\cite{bertiniExactCorrelationFunctions2019}.} In the
$(q, r) = (3, 1)$ case, we find that two out of three
one-site operators ${\rm s}^\zz$ which constitute the coarse-grained ${\rm s}^\zz(x, t)$ [see Eq.~\eqref{eq:coarse-grained-spin}] move east with the average velocity $\frac{1}{2}$ (i.e., for $\frac{3}{2}$ sites per time period, on average). One of them instead moves west with velocity $1$ (i.e., for $3$ sites per time period). This is depicted in
Fig.~\ref{fig:perm}. The contributions to the first moment of $S(x, t)$ thus cancel out. However, as we will see in the following section, this is not the case for the higher odd moments.

\begin{figure}[ht!]
\centering
\includegraphics[width=0.8\textwidth]{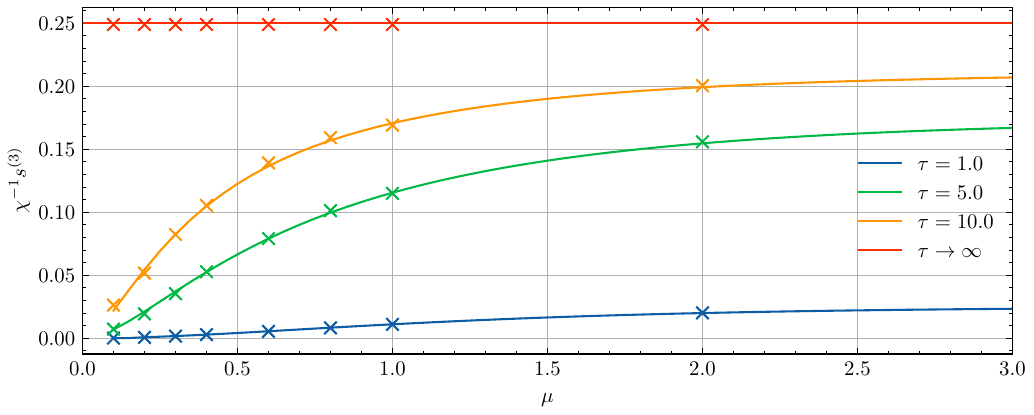}
\includegraphics[width=0.8\textwidth]{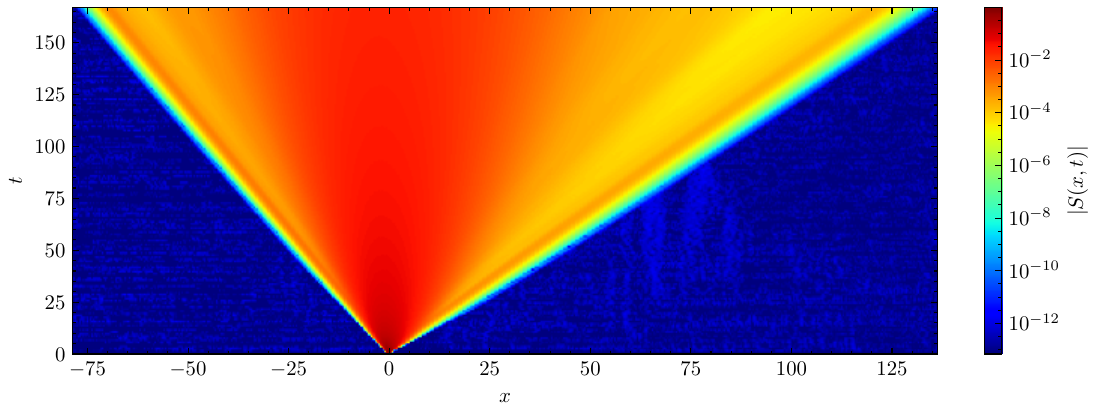}
\caption{The top panel shows numerically obtained normalized third moment $\chi^{-1}s^{(3)}$ of the dynamical spin susceptibility (crosses), compared with the analytical prediction. The bottom panel shows the dynamical spin susceptibility for $\tau = 1$ and $\mu = 0.2$. Calculations were performed in the circuit with isotropic Heisenberg gates arranged in the asymmetric $(q, r) = (3, 1)$ geometry shown in Fig.~\hyperref[fig:circuit-q3r1]{3(b)}. The circuit acts on $L = 1002$ qubits up to times $t = 167$. We have used the linear-response approximation~\eqref{eq:lin_resp} with $\Delta \mu = 10^{-4}$. The maximal bond dimension in the TEBD algorithm was set to $d_{\rm max} = 256$. The third moment is determined by fitting a linear function to $(\chi^{-1}s^{(3)})^{1/3}t$.}
\label{fig:numerics1}
\end{figure}

\subsubsection{Asymmetry without a drift.}

While the unitary gates are symmetric, their local arrangement can still break the space-reflection symmetry $\mathcal{P}$. We therefore expect some higher moments $s^{(n)}$, for odd $n$, to be nonzero. Indeed, the nonzero hydrodynamic prediction for the normalized third moment of the dynamical susceptibility, $\chi^{-1}s^{(3)}$, is corroborated by extensive numerical simulations---see Fig.~\ref{fig:numerics1}. 

Numerical simulations are performed using the TEBD algorithm. We compute the dynamical spin susceptibility by invoking the linear-response approximation (see Ref.~\cite{ljubotinaKardarparisizhangPhysicsQuantum2019})
\begin{equation}
    S(x, t) \approx \frac{\langle {\rm s}^\zz(x - 1,t) \rangle_{\mu, \Delta \mu} - \langle {\rm s}^\zz(x,t) \rangle_{\mu, \Delta \mu}}{\Delta \mu},
    \label{eq:lin_resp}
\end{equation}
which is valid for small $\Delta \mu$. Here, $\langle \bullet \rangle_{\mu, \Delta \mu}$ denotes the expectation value in the weakly polarized domain-wall state 
\begin{equation}
    \varrho_{\mu, \Delta \mu} = \frac{1}{Z_{\mu,\Delta\mu}}\e^{(-\mu + \Delta \mu) \sum_{x\le 0} {\rm s}^\zz(x)} \e^{(-\mu - \Delta \mu) \sum_{x'>0}{\rm s}^\zz(x')},
\end{equation}
$Z_{\mu,\Delta\mu}$ being chosen so that the state is normalized, ${\rm Tr}(\varrho_{\mu, \Delta \mu})=1$.

\section{Conclusion}
\label{sec:conclusion}

We have investigated integrable quantum circuit geometries that generalize the brickwork-type integrable Trotterization of a continuous-time dynamics. We have explicitly demonstrated the integrability of all circuits in which a two-qubit unitary gate satisfying the Yang-Baxter equation is applied to each pair of neighbouring qubits exactly once per time period. Additionally, we have described a systematic approach for identifying commuting transfer matrices that generate such circuits. Our construction is not limited to the same-gate circuits: it can also be used to build integrable circuits in which the gate parameter changes in discrete time.

Besides the brickwork-type Trotterization of an integrable continuous-time dynamical system, there are other integrable circuit geometries which utilize the same unitary gate but may exhibit qualitatively different large-scale dynamics. This suggests that the dynamics in discrete spacetime can be inherently richer than the one governed by a many-body Hamiltonian in continuous time. We have investigated the features of a discrete-spacetime dynamics by studying spin transport in the circuits constructed using the U(1)-invariant isotropic Heisenberg gate. Using hydrodynamic description of dynamical correlation functions, supported by extensive TEBD simulations, we found that a spatially asymmetric arrangement of otherwise symmetric Heisenberg gates induces an asymmetry in the dynamical spin susceptibility. Specifically, the latter's odd moments of third or higher order become nonzero. However, the first (noncentered) moment, which corresponds to a drift in the correlation spreading, remains exactly zero. We demonstrated that this is due to the cancellation of charge contributions carried by identical quasiparticles---a condition that holds as long as the unitary gates act on pairs of identical degrees of freedom (e.g., pairs of qubits).

Our work raises several interesting questions. One concerns the identification of classical counterparts to our circuits, which would extend the framework of classical integrable Trotterizations~\cite{krajnik2020}. A related open question is whether integrable classical circuits with multiple layers of distinct local symplectic maps (replacing the unitary gates) relate to the integrable fishnet stochastic circuits proposed in Ref.\cite{krajnik2024}. Additionally, the existence of integrable circuit geometries involving multi-qubit gates\cite{gomborIntegrableSpinChains2021,vona2024} or unitary gates acting on distinct spins or other degrees of freedom~\cite{zadnikQuantumManybodySpin2024a,richelli2024} should be explored. Here, it would be interesting to investigate the competing effects of an asymmetric gate arrangement on the one side, and chiral propagation of different degrees of freedom on the other. Another possible future direction concerns circuit geometries with open boundary conditions, as well as integrable dissipative and nonunitary circuits~\cite{vanicat2018,sa2021,sumartin,vernier2024,paletta2025,popkov2025} (see Refs.~\cite{medvedyeva2016,buca2020,ziolkowska2020,essler2020,deLeeuw2021,deLeeuw2024} for some examples of integrable dissipative systems in continuous time). Finally, we cannot avoid emphasizing that a wide range of integrable quantum circuit geometries, composed using an arbitrary U(1)-invariant two-qubit unitary gate~\cite{znidaricIntegrabilityGenericHomogeneous2024}, provides ample opportunities for experimental investigation of unconventional symmetries and phases of matter~\cite{znidaricInhomogeneousSU2Symmetries2024} using modern quantum computing platforms.

\section*{Acknowledgements}

The Authors thank Marko \v{Z}nidari\v{c} and Yuan Miao for insightful discussions. This work has been supported by: Slovenian Research Agency (ARIS) under Research Startup Program No. SN-ZRD/22-27/0510---NESY (L.Z.), and Research Program P1-0402 (L.Z. and U.D.); European Research Council (ERC) under Advanced Grant No. 101096208---QUEST (L.Z.); European Union HORIZON CL4-2022-QUANTUM-02 SGA through PASQuanS2.1 Grant Agreement No. 101113690 (C.P.); Hungarian National Research, Development and Innovation Office, NKFIH Excellence Grant TKP2021-NKTA-64 (B.P.). Numerical simulations were performed using the ITensor library~\cite{fishmanItensorSoftwareLibrary2022}.

\section*{References}
\bibliographystyle{iopart-num}
\bibliography{references.bib}

\providecommand{\newblock}{}
\begin{thebibliography}{10}
\expandafter\ifx\csname url\endcsname\relax
  \def\url#1{{\tt #1}}\fi
\expandafter\ifx\csname urlprefix\endcsname\relax\def\urlprefix{URL }\fi
\providecommand{\eprint}[2][]{\url{#2}}

\bibitem{baxter1982}
Baxter R~J 1982 {\em Exactly solved models in statistical mechanics\/} (London:
  Academic Press)

\bibitem{calabrese2016}
Calabrese P, Essler F~H~L and Mussardo G 2016 Introduction to ‘{Q}uantum
  integrability in out of equilibrium systems’ {\em J. Stat. Mech. - Theory
  E.\/} \href{http://dx.doi.org/10.1088/1742-5468/2016/06/064001}{{\bf 2016}
  064001}

\bibitem{faddeev1996}
Faddeev L~D 1996 How {A}lgebraic {B}ethe {A}nsatz works for integrable model
  (\textit{Preprint} \eprint{hep-th/9605187})

\bibitem{destri1987}
Destri C and {De Vega} H 1987 Light-cone lattice approach to fermionic theories
  in 2{D}: {T}he massive {T}hirring model {\em Nucl. Phys. B\/}
  \href{http://dx.doi.org/https://doi.org/10.1016/0550-3213(87)90193-3}{{\bf
  290} 363--391}

\bibitem{faddeev1994}
Faddeev L and Volkov A~Y 1994 Hirota equation as an example of an integrable
  symplectic map {\em Lett. Math. Phys.\/}
  \href{http://dx.doi.org/10.1007/BF00739422}{{\bf 32} 125--135}

\bibitem{gritsev2017}
Gritsev V and Polkovnikov A 2017 {Integrable Floquet dynamics} {\em SciPost
  Phys.\/} \href{http://dx.doi.org/10.21468/SciPostPhys.2.3.021}{{\bf 2} 021}

\bibitem{vanicat2018}
Vanicat M, Zadnik L and Prosen T 2018 Integrable {T}rotterization: Local
  conservation laws and boundary driving {\em Phys. Rev. Lett.\/}
  \href{http://dx.doi.org/10.1103/PhysRevLett.121.030606}{{\bf 121}(3) 030606}

\bibitem{vanicat2018integrable}
Vanicat M 2018 Integrable {F}loquet dynamics, generalized exclusion processes
  and “fused” matrix ansatz {\em Nucl. Phys. B\/}
  \href{http://dx.doi.org/10.1016/j.nuclphysb.2018.02.007}{{\bf 929} 298--329}

\bibitem{suzukiGeneralizedTrottersFormula1976}
Suzuki M 1976 Generalized {Trotter}'s formula and systematic approximants of
  exponential operators and inner derivations with applications to many-body
  problems {\em Commun. Math. Phys.\/}
  \href{http://dx.doi.org/10.1007/BF01609348}{{\bf 51} 183--190}

\bibitem{kauffman2004}
Kauffman L~H and Lomonaco S~J 2004 Braiding operators are universal quantum
  gates {\em New J. Phys.\/}
  \href{http://dx.doi.org/10.1088/1367-2630/6/1/134}{{\bf 6} 134}

\bibitem{zhang2005}
Zhang Y, Kauffman L~H and Ge M~L 2005 Universal quantum gate,
  {Y}ang–{B}axterization and {H}amiltonian {\em Int. J. Quantum Inf.\/}
  \href{http://dx.doi.org/10.1142/S0219749905001547}{{\bf 03} 669--678}

\bibitem{zhang2024geometric}
Zhang K, Hao K, Yu K, Korepin V and Yang W~L 2024 Geometric representations of
  braid and {Y}ang–{B}axter gates {\em J. Phys. A - Math. Theor.\/}
  \href{http://dx.doi.org/10.1088/1751-8121/ad85b2}{{\bf 57} 445303}

\bibitem{zhang2024optimal}
Zhang K, Yu K, Hao K and Korepin V 2024 Optimal realization of
  {Y}ang–{B}axter gate on quantum computers {\em Advanced Quantum
  Technologies\/}
  \href{http://dx.doi.org/https://doi.org/10.1002/qute.202300345}{{\bf 7}
  2300345}

\bibitem{ljubotinaBallisticSpinTransport2019}
Ljubotina M, Zadnik L and Prosen T 2019 Ballistic spin transport in a
  periodically driven integrable quantum system {\em Phys. Rev. Lett.\/}
  \href{http://dx.doi.org/10.1103/PhysRevLett.122.150605}{{\bf 122} 150605}

\bibitem{ljubotinaKardarparisizhangPhysicsQuantum2019}
Ljubotina M, Žnidarič M and Prosen T 2019 Kardar-{P}arisi-{Z}hang physics in
  the quantum {H}eisenberg magnet {\em Phys. Rev. Lett.\/}
  \href{http://dx.doi.org/10.1103/PhysRevLett.122.210602}{{\bf 122} 210602}

\bibitem{rosenbergDynamicsMagnetizationInfinite2024}
Rosenberg E {\em et~al.\/} 2024 Dynamics of magnetization at infinite
  temperature in a {Heisenberg} spin chain {\em Science\/}
  \href{http://dx.doi.org/10.1126/science.adi7877}{{\bf 384} 48--53}

\bibitem{summer2024}
Summer A, Nico-Katz A, Dooley S and Goold J 2024 Anomalous transport in
  {U}(1)-symmetric quantum circuits (\textit{Preprint} \eprint{2411.14357})

\bibitem{aleiner2021}
Aleiner I~L 2021 {B}ethe {A}nsatz solutions for certain periodic quantum
  circuits {\em Ann. Phys.\/}
  \href{http://dx.doi.org/https://doi.org/10.1016/j.aop.2021.168593}{{\bf 433}
  168593}

\bibitem{maruyoshi2023}
Maruyoshi K, Okuda T, Pedersen J~W, Suzuki R, Yamazaki M and Yoshida Y 2023
  Conserved charges in the quantum simulation of integrable spin chains {\em J.
  Phys. A - Math. Theor.\/}
  \href{http://dx.doi.org/10.1088/1751-8121/acc369}{{\bf 56} 165301}

\bibitem{hillberry2024}
Hillberry L~E, Piroli L, Vernier E, Halpern N~Y, Prosen T and Carr L~D 2024
  Integrability of {G}oldilocks quantum cellular automata (\textit{Preprint}
  \eprint{2404.02994})

\bibitem{hutsalyuk2024}
Hutsalyuk A, Jiang Y, Pozsgay B, Xu H and Zhang Y 2024 Exact spin correlators
  of integrable quantum circuits from algebraic geometry (\textit{Preprint}
  \eprint{2405.16070})

\bibitem{morvan2022}
Morvan A {\em et~al.\/} 2022 Formation of robust bound states of interacting
  microwave photons {\em Nature\/}
  \href{http://dx.doi.org/10.1038/s41586-022-05348-y}{{\bf 612} 240–245}

\bibitem{hudomal2024}
Hudomal A, Smith R, Hallam A and Papi\ifmmode~\acute{c}\else \'{c}\fi{} Z 2024
  Integrability breaking and bound states in {G}oogle's decorated {XXZ}
  circuits {\em PRX Quantum\/}
  \href{http://dx.doi.org/10.1103/PRXQuantum.5.010316}{{\bf 5}(1) 010316}

\bibitem{surace2024}
Surace F~M and Motrunich O 2024 Robustness and eventual slow decay of bound
  states of interacting microwave photons in the {G}oogle {Q}uantum {AI}
  experiment {\em PRX Quantum\/}
  \href{http://dx.doi.org/10.1103/PRXQuantum.5.010317}{{\bf 5}(1) 010317}

\bibitem{znidaricInhomogeneousSU2Symmetries2024}
Žnidarič M 2024 Inhomogeneous {SU}(2) symmetries in homogeneous integrable
  {U}(1) circuits and transport (\textit{Preprint} \eprint{2412.09371})

\bibitem{vernier2024}
Vernier E, Yeh H~C, Piroli L and Mitra A 2024 Strong zero modes in integrable
  quantum circuits {\em Phys. Rev. Lett.\/}
  \href{http://dx.doi.org/10.1103/PhysRevLett.133.050606}{{\bf 133}(5) 050606}

\bibitem{vernier2023}
Vernier E, Bertini B, Giudici G and Piroli L 2023 Integrable digital quantum
  simulation: {G}eneralized {G}ibbs ensembles and {T}rotter transitions {\em
  Phys. Rev. Lett.\/}
  \href{http://dx.doi.org/10.1103/PhysRevLett.130.260401}{{\bf 130}(26) 260401}

\bibitem{znidaricIntegrabilityGenericHomogeneous2024}
Žnidaric M, Duh U and Zadnik L 2024 Integrability is generic in homogeneous
  {U}(1)-invariant nearest-neighbor qubit circuits (\textit{Preprint}
  \eprint{2410.06760})

\bibitem{suchsland2025}
Suchsland P, Moessner R and Claeys P~W 2025 Krylov complexity and {T}rotter
  transitions in unitary circuit dynamics {\em Phys. Rev. B\/}
  \href{http://dx.doi.org/10.1103/PhysRevB.111.014309}{{\bf 111}(1) 014309}

\bibitem{friedman2019}
Friedman A~J, Chan A, De~Luca A and Chalker J~T 2019 Spectral statistics and
  many-body quantum chaos with conserved charge {\em Phys. Rev. Lett.\/}
  \href{http://dx.doi.org/10.1103/PhysRevLett.123.210603}{{\bf 123}(21) 210603}

\bibitem{sa2021}
S\'a L, Ribeiro P and Prosen T 2021 Integrable nonunitary open quantum circuits
  {\em Phys. Rev. B\/}
  \href{http://dx.doi.org/10.1103/PhysRevB.103.115132}{{\bf 103}(11) 115132}

\bibitem{paletta2025}
Paletta C and Prosen T 2025 {Integrability of open boundary driven quantum
  circuits} {\em SciPost Phys.\/}
  \href{http://dx.doi.org/10.21468/SciPostPhys.18.1.027}{{\bf 18} 027}

\bibitem{popkov2025}
Popkov V and Prosen T 2025 Exact {NESS} of {XXZ} circuits boundary driven with
  arbitrary resets or fields (\textit{Preprint} \eprint{2502.06731})

\bibitem{gombor2022}
Gombor T and Pozsgay B 2022 {Superintegrable cellular automata and dual unitary
  gates from Yang-Baxter maps} {\em SciPost Phys.\/}
  \href{http://dx.doi.org/10.21468/SciPostPhys.12.3.102}{{\bf 12} 102}

\bibitem{gombor2024}
Gombor T and Pozsgay B 2024 {Integrable deformations of superintegrable quantum
  circuits} {\em SciPost Phys.\/}
  \href{http://dx.doi.org/10.21468/SciPostPhys.16.4.114}{{\bf 16} 114}

\bibitem{singh2024}
Singh V~K, Sinha A, Padmanabhan P and Korepin V 2024 Unitary tetrahedron
  quantum gates (\textit{Preprint} \eprint{2407.10731})

\bibitem{sinha2024}
Sinha A, Padmanabhan P and Korepin V 2024 Toffoli gates solve the tetrahedron
  equations (\textit{Preprint} \eprint{2405.16477})

\bibitem{castro-alvaredoEmergentHydrodynamicsIntegrable2016}
Castro-Alvaredo O~A, Doyon B and Yoshimura T 2016 Emergent hydrodynamics in
  integrable quantum systems out of equilibrium {\em Phys. Rev. X\/}
  \href{http://dx.doi.org/10.1103/PhysRevX.6.041065}{{\bf 6} 041065}

\bibitem{bertiniTransportOutofequilibriumXXZ2016}
Bertini B, Collura M, De~Nardis J and Fagotti M 2016 Transport in
  out-of-equilibrium {XXZ} chains: {Exact} profiles of charges and currents
  {\em Phys. Rev. Lett.\/}
  \href{http://dx.doi.org/10.1103/PhysRevLett.117.207201}{{\bf 117} 207201}

\bibitem{doyon2020}
Doyon B 2020 Lecture notes on generalised hydrodynamics {\em SciPost Phys.
  Lect. Notes\/} \href{http://dx.doi.org/10.21468/SciPostPhysLectNotes.18}{ 18}

\bibitem{hubner2025}
Hübner F, Vernier E and Piroli L 2025 Generalized hydrodynamics of integrable
  quantum circuits (\textit{Preprint} \eprint{2408.00474})

\bibitem{zauner2015}
Zauner V, Ganahl M, Evertz H~G and Nishino T 2015 Time evolution within a
  comoving window: scaling of signal fronts and magnetization plateaus after a
  local quench in quantum spin chains {\em J. Phys. - Condens. Mat.\/}
  \href{http://dx.doi.org/10.1088/0953-8984/27/42/425602}{{\bf 27} 425602}

\bibitem{bertini2016}
Bertini B and Fagotti M 2016 Determination of the nonequilibrium steady state
  emerging from a defect {\em Phys. Rev. Lett.\/}
  \href{http://dx.doi.org/10.1103/PhysRevLett.117.130402}{{\bf 117}(13) 130402}

\bibitem{eisler2020}
Eisler V and Maislinger F 2020 {Front dynamics in the XY chain after local
  excitations} {\em SciPost Phys.\/}
  \href{http://dx.doi.org/10.21468/SciPostPhys.8.3.037}{{\bf 8} 037}

\bibitem{gruber2021}
Gruber M and Eisler V 2021 {Entanglement spreading after local fermionic
  excitations in the XXZ chain} {\em SciPost Phys.\/}
  \href{http://dx.doi.org/10.21468/SciPostPhys.10.1.005}{{\bf 10} 005}

\bibitem{bidzhiev2022}
Bidzhiev K, Fagotti M and Zadnik L 2022 Macroscopic effects of localized
  measurements in jammed states of quantum spin chains {\em Phys. Rev. Lett.\/}
  \href{http://dx.doi.org/10.1103/PhysRevLett.128.130603}{{\bf 128}(13) 130603}

\bibitem{fagotti2022}
Fagotti M 2022 Global quenches after localized perturbations {\em Phys. Rev.
  Lett.\/} \href{http://dx.doi.org/10.1103/PhysRevLett.128.110602}{{\bf
  128}(11) 110602}

\bibitem{zadnik2022}
Zadnik L, Bocini S, Bidzhiev K and Fagotti M 2022 Measurement catastrophe and
  ballistic spread of charge density with vanishing current {\em J. Phys. A -
  Math. Theor.\/} \href{http://dx.doi.org/10.1088/1751-8121/aca254}{{\bf 55}
  474001}

\bibitem{zadnikQuantumManybodySpin2024a}
Zadnik L, Ljubotina M, Krajnik v, Ilievski E and Prosen T 2024 Quantum
  many-body spin ratchets {\em PRX Quantum\/}
  \href{http://dx.doi.org/10.1103/PRXQuantum.5.030356}{{\bf 5} 030356}

\bibitem{richelli2024}
Richelli P, Schoutens K and Zorzato A 2024 {Brick wall quantum circuits with
  global fermionic symmetry} {\em SciPost Phys.\/}
  \href{http://dx.doi.org/10.21468/SciPostPhys.17.3.087}{{\bf 17} 087}

\bibitem{gong2022}
Gong Z, Nahum A and Piroli L 2022 Coarse-grained entanglement and operator
  growth in anomalous dynamics {\em Phys. Rev. Lett.\/}
  \href{http://dx.doi.org/10.1103/PhysRevLett.128.080602}{{\bf 128}(8) 080602}

\bibitem{duhClassificationSamegateQuantum2024}
Duh U and Žnidarič M 2024 Classification of same-gate quantum circuits and
  their space-time symmetries with application to the level-spacing
  distribution {\em Phys. Rev. Res.\/}
  \href{http://dx.doi.org/10.1103/PhysRevResearch.6.023068}{{\bf 6} 023068}

\bibitem{sogo1982}
Sogo K, Uchinami M, Akutsu Y and Wadati M 1982 Classification of exactly
  solvable two-component models {\em Prog. Theor. Phys.\/}
  \href{http://dx.doi.org/10.1143/PTP.68.508}{{\bf 68} 508--526}

\bibitem{beisert2013}
Beisert N, Fiévet L, de~Leeuw M and Loebbert F 2013 Integrable deformations of
  the {XXZ} spin chain {\em J. Stat. Mech. - Theory E.\/}
  \href{http://dx.doi.org/10.1088/1742-5468/2013/09/P09028}{{\bf 2013} P09028}

\bibitem{vieira2018}
Vieira R~S 2018 Solving and classifying the solutions of the {Y}ang-{B}axter
  equation through a differential approach. {T}wo-state systems {\em J. High
  Energy Phys.\/} \href{http://dx.doi.org/10.1007/JHEP10(2018)110}{{\bf 2018}
  110}

\bibitem{deLeeuw2020}
de~Leeuw M, Paletta C, Pribytok A, Retore A~L and Ryan P 2020 Classifying
  nearest-neighbor interactions and deformations of {AdS} {\em Phys. Rev.
  Lett.\/} \href{http://dx.doi.org/10.1103/PhysRevLett.125.031604}{{\bf 125}(3)
  031604}

\bibitem{yangbaxterboostdeleeuw}
de~Leeuw M, Paletta C, Pribytok A, Retore A~L and Ryan P 2021 {Y}ang-{B}axter
  and the boost: {S}plitting the difference {\em SciPost Phys.\/}
  \href{http://dx.doi.org/10.21468/SciPostPhys.11.3.069}{{\bf 11} 069}

\bibitem{miaoFloquetBaxterisation2024}
Miao Y, Gritsev V and Kurlov D~V 2024 The floquet baxterisation {\em SciPost
  Phys.\/} \href{http://dx.doi.org/10.21468/SciPostPhys.16.3.078}{{\bf 16} 078}

\bibitem{difrancesco2006}
Francesco P~D and Zinn-Justin P 2006 Around the {R}azumov-{S}troganov
  conjecture: proof of a multi-parameter sum rule (\textit{Preprint}
  \eprint{math-ph/0410061})

\bibitem{baxter97}
Baxter R~J 1978 Solvable eight-vertex model on an arbitrary planar lattice {\em
  Philos. T. R. Soc. A\/} \href{http://dx.doi.org/10.1098/rsta.1978.0062}{{\bf
  289} 315--346}

\bibitem{krajnik2024}
Žiga Krajnik, Ilievski E, Prosen T, Héry B~J~A and Pasquier V 2024 Integrable
  fishnet circuits and {B}rownian solitons (\textit{Preprint}
  \eprint{2411.08030})

\bibitem{gomborIntegrableSpinChains2021}
Gombor T and Pozsgay B 2021 Integrable spin chains and cellular automata with
  medium-range interaction {\em Phys. Rev. E\/}
  \href{http://dx.doi.org/10.1103/PhysRevE.104.054123}{{\bf 104} 054123}

\bibitem{takahashi1999}
Takahashi M 1999 {\em Thermodynamics of One-Dimensional Solvable Models\/}
  (Cambridge University Press)

\bibitem{yang1969}
Yang C~N and Yang C~P 1969 Thermodynamics of a one‐dimensional system of
  bosons with repulsive delta‐function interaction {\em J. Math. Phys.\/}
  \href{http://dx.doi.org/10.1063/1.1664947}{{\bf 10} 1115--1122}

\bibitem{ilievski2018}
Ilievski E, De~Nardis J, Medenjak M and Prosen T 2018 Superdiffusion in
  one-dimensional quantum lattice models {\em Phys. Rev. Lett.\/}
  \href{http://dx.doi.org/10.1103/PhysRevLett.121.230602}{{\bf 121}(23) 230602}

\bibitem{doyon2017}
Doyon B and Spohn H 2017 Drude weight for the {L}ieb-{L}iniger {B}ose gas {\em
  SciPost Phys.\/} \href{http://dx.doi.org/10.21468/SciPostPhys.3.6.039}{{\bf
  3} 039}

\bibitem{denardis2019}
Nardis J~D, Bernard D and Doyon B 2019 {Diffusion in generalized hydrodynamics
  and quasiparticle scattering} {\em SciPost Phys.\/}
  \href{http://dx.doi.org/10.21468/SciPostPhys.6.4.049}{{\bf 6} 049}

\bibitem{denardis2022}
De~Nardis J, Doyon B, Medenjak M and Panfil M 2022 Correlation functions and
  transport coefficients in generalised hydrodynamics {\em J. Stat. Mech. -
  Theory E.\/} \href{http://dx.doi.org/10.1088/1742-5468/ac3658}{{\bf 2022}
  014002}

\bibitem{ilievski2021}
Ilievski E, De~Nardis J, Gopalakrishnan S, Vasseur R and Ware B 2021
  Superuniversality of superdiffusion {\em Phys. Rev. X\/}
  \href{http://dx.doi.org/10.1103/PhysRevX.11.031023}{{\bf 11}(3) 031023}

\bibitem{bertiniExactCorrelationFunctions2019}
Bertini B, Kos P and Prosen T 2019 Exact correlation functions for dual-unitary
  lattice models in 1+1 dimensions {\em Phys. Rev. Lett.\/}
  \href{http://dx.doi.org/10.1103/PhysRevLett.123.210601}{{\bf 123} 210601}

\bibitem{krajnik2020}
Žiga Krajnik, Ilievski E and Prosen T 2020 {Integrable matrix models in
  discrete space-time} {\em SciPost Phys.\/}
  \href{http://dx.doi.org/10.21468/SciPostPhys.9.3.038}{{\bf 9} 038}

\bibitem{vona2024}
Vona I, Mestyán M and Pozsgay B 2024 Exact real time dynamics with free
  fermions in disguise (\textit{Preprint} \eprint{2405.20832})

\bibitem{sumartin}
Su L and Martin I 2022 Integrable nonunitary quantum circuits {\em Phys. Rev.
  B\/} \href{http://dx.doi.org/10.1103/PhysRevB.106.134312}{{\bf 106}(13)
  134312}

\bibitem{medvedyeva2016}
Medvedyeva M~V, Essler F~H~L and Prosen T 2016 Exact bethe ansatz spectrum of a
  tight-binding chain with dephasing noise {\em Phys. Rev. Lett.\/}
  \href{http://dx.doi.org/10.1103/PhysRevLett.117.137202}{{\bf 117}(13) 137202}

\bibitem{buca2020}
Buča B, Booker C, Medenjak M and Jaksch D 2020 Bethe ansatz approach for
  dissipation: exact solutions of quantum many-body dynamics under loss {\em
  New Journal of Physics\/}
  \href{http://dx.doi.org/10.1088/1367-2630/abd124}{{\bf 22} 123040}

\bibitem{ziolkowska2020}
Ziolkowska A~A and Essler F~H 2020 {Yang-Baxter integrable Lindblad equations}
  {\em SciPost Phys.\/}
  \href{http://dx.doi.org/10.21468/SciPostPhys.8.3.044}{{\bf 8} 044}

\bibitem{essler2020}
Essler F~H~L and Piroli L 2020 Integrability of one-dimensional lindbladians
  from operator-space fragmentation {\em Phys. Rev. E\/}
  \href{http://dx.doi.org/10.1103/PhysRevE.102.062210}{{\bf 102}(6) 062210}

\bibitem{deLeeuw2021}
de~Leeuw M, Paletta C and Pozsgay B 2021 Constructing integrable lindblad
  superoperators {\em Phys. Rev. Lett.\/}
  \href{http://dx.doi.org/10.1103/PhysRevLett.126.240403}{{\bf 126}(24) 240403}

\bibitem{deLeeuw2024}
de~Leeuw M, Paletta C, Pozsgay B and Vernier E 2024 Hidden quasilocal charges
  and gibbs ensemble in a lindblad system {\em Phys. Rev. B\/}
  \href{http://dx.doi.org/10.1103/PhysRevB.109.054311}{{\bf 109}(5) 054311}

\bibitem{fishmanItensorSoftwareLibrary2022}
Fishman M, White S~R and Stoudenmire E~M 2022 The {IT}ensor software library
  for tensor network calculations {\em SciPost Phys. Codebases\/}
  \href{http://dx.doi.org/10.21468/SciPostPhysCodeb.4}{ 4}

\end{thebibliography}

\end{document}